\documentclass{aa}

\usepackage{psfrag,color}
\usepackage{amsmath}
\usepackage{amssymb}
\usepackage{textcomp}
\usepackage{txfonts}
\usepackage{natbib}
\usepackage{graphicx}
\usepackage{rotating}
\usepackage{xspace}
\usepackage{soul}
\bibpunct{(}{)}{;}{a}{}{,}

\begin{document}

\title{Extended object reconstruction in adaptive-optics imaging: the multiresolution approach}

\author{Roberto Baena Gall\'e \inst{\ref{ub},{\ref{racab}}}\thanks{\email{rbaena@am.ub.es}}
\and{Jorge N\'u\~nez \inst{\ref{ub},{\ref{racab}}}}\thanks{\email{jorge@am.ub.es}}
\and{Szymon Gladysz \inst{\ref{fraunhofer}}}
}

\institute{Departament d'Astronomia i Meteorologia i Institut de Ci\`encies del Cosmos (ICC), Universitat de Barcelona (UB/IEEC), Mart\'i i Franqu\'es 1, E-08025 Barcelona, Spain.\label{ub}
\and
Observatori Fabra, Reial Acad\`emia de Ci\`encies i Arts de Barcelona, Cam\'i de l'Observatori s/n, E-08002 Barcelona, Spain.\label{racab}
\and
Fraunhofer-Institut f\"{u}r Optronik, Systemtechnik und Bildauswertung. Gutleuthausstra\ss e 1. 76275 Ettlingen, Germany.
\newline
 \email{szymon.gladysz@iosb.fraunhofer.de}\label{fraunhofer}
}

\date{Received 00 xxxxxx 2011 / Accepted 00 xxxxxx 2011}

\abstract
{}{We propose the application of multiresolution transforms, such as wavelets (WT) and curvelets (CT), to the reconstruction of images of extended objects that have been acquired with adaptive optics (AO) systems. Such multichannel approaches normally make use of probabilistic tools in order to distinguish significant structures from noise and reconstruction residuals. Therefore, we also test the performance of two different probabilistic masks: one based on local correlation and the other on local standard deviation. Furthermore, we aim to check the historical assumption that image-reconstruction algorithms using static PSFs are not suitable for AO imaging.}
{We convolve an image of Saturn taken with the Hubble Space Telescope (HST) with AO PSFs from the 5-m Hale telescope at the Palomar Observatory and add both shot and readout noise. Subsequently, we apply different approaches to the blurred and noisy data in order to recover the original object. The approaches include multi-frame blind deconvolution (with the algorithm IDAC), myopic deconvolution with regularization (with MISTRAL) and wavelets- or curvelets-based static PSF deconvolution ({AWMLE and ACMLE algorithms}). We used {the mean squared error (MSE) and the} structural similarity index (SSIM), also known as the Wang-Bovik index, to compare the results. {We discuss the strengths and weaknesses of the two metrics and the usefulness of using more than one metric to evaluate the results of imaging restoration.}}
{{We found that CT produces better results than WT, as measured in terms of MSE and SSIM. In the wavelet domain,} the mask based on local correlation provides better results than the mask based on standard deviation for the same number of iterations. Furthermore, multichannel deconvolution with a static PSF produces results which are generally better than the results obtained with the myopic/blind approaches (for the images we tested) thus showing that the ability of a method to suppress the noise and to track the underlying iterative process is just as critical as the capability of the myopic/blind approaches to update the PSF.}{}

\keywords{Instrumentation: adaptive optics - Techniques: image processing - Methods: miscellaneous }

\titlerunning{Extended object reconstruction in AO: the multiresolution approach.}
\authorrunning{R.Baena Gall\'e et al.}
\maketitle

\section{INTRODUCTION}

The distortions introduced into images by the acquisition process in astronomical ground-based observations are well known. Apart from the most common, such as vignetting, non-zero background or bad pixels, which must be removed prior to any other analysis, atmospheric turbulence limits the spatial resolution of an image whereas the electronic devices used to acquire and amplify the signal introduces noise. An image is also corrupted by Poisson noise due to fluctuations in the number of received photons at each pixel \citep{Andrews1977}. The classical equation that describes the image formation process is:

{ }
\begin{equation}
 image = [PSF \ast object] \Diamond noise
\label{eq1}
\end{equation}
{ }

where $\ast$ denotes the convolution operation. The symbol $\Diamond$ is a pixel-by-pixel operation which reduces to the simple addition in the case when noise is additive and independent of $[PSF \ast object]$, while for Poisson noise it is an operation which returns a random deviate drawn from a Poisson distribution with mean equal to $[PSF \ast object]$. It is well known that direct inversion of equation \ref{eq1} in the Fourier domain amplifies noisy frequencies close to the cut-off frequency. Hence, in the presence of noise, such a simple method cannot be used.

Several deconvolution approaches have been proposed in order to estimate the original signal from the seeing-limited and noise-degraded data. Since equation \ref{eq1} is an ill-posed problem, with non-unique stable solutions, one approach is to regularize the Fourier inversion in order to constrain possible solutions \citep{Tikhonov1987,Bertero1998}. This method generally imposes a trade-off between noise amplification and the desired resolution which generally leads to smooth solutions. Bayesian methodology \citep{Molina2001,Starck2002b} allows a solution compatible with the statistical nature of the signal to be sought, leading to maximum likelihood estimators (MLE) \citep{Richardson1972,Lucy1974} or maximum a posteriori (MAP) approaches if prior information is used, e.g., the positivity of the signal or entropy \citep{Frieden1978,Jaynes1982}.

All these methods can be enhanced through multichannel analysis by decomposing the signal in different planes, each of them representative of a certain scale of resolution. In such a decomposition, fine details in an image are confined to some planes, whereas coarse structures are confined to others. One of the most powerful ways to perform such decomposition is by means of the wavelet transform (WT). {In particular in the astronomical context, the undecimated isotropic \`a trous algorithm} \citep{Holschneider1989,Shensa1992,Starck1994}, {also known as the Starlet transform, is often used}. The WT creates a multiple representation of a signal, classifying its frequencies and, simultaneously, spatially localizing them in the field of view. {For the specific case of the Starlet transfrom}, this can be expressed by:

\begin{equation}
 \mathbf{s=\omega_0^s + \omega_1^s + \omega_2^s +...+ \omega_N^s + r_N^s }
\label{eq5}
\end{equation}

where $\mathbf{s}$ is the signal being decomposed into wavelet coefficients, $\mathbf{\omega_j^s}$ is the wavelet plane at resolution \textit{j} and $\mathbf{r_N^s}$ is the residual wavelet plane. {We point out that equation} \ref{eq5} {provides a prescription for direct reconstruction of the original image from all of the wavelet planes and the residual plane}. The advantatge of WT is that it allows for different strategies to be used for different wavelet planes, e.g., by defining thresholds of statistical significance.

Given the advantages of the multiresolution analysis the aforementioned deconvolution approaches have been adapted to work in the wavelet domain. The maximum-likelihood method (MLE) was modified into a two-channel algorithm \citep{Lucy1994} where the first channel corresponds to the signal contribution and the second to the background. The wavelet transform has also been applied to maximum-entropy deconvolution methods \citep{Nunez1998} in order to segment the image and apply different regularization parameters to each region. \citet{Starck2001} generalizes the method of maximum-entropy within a wavelet framework, separating the problem into two stages: noise control in the image domain and smoothness in the object domain.

While WT has been widely used in astronomical image analysis and in deconvolution, the reported use (in the same context) of multi-transforms with properties that improve on or complement WT, is scarce. Such methods include, among many others: waveatoms \citep{Demanet2007}, which aim to represent signals by textures; and curvelets \citep{Candes2006}, which introduce orientation as a classification parameter together with frequency and position. \citet{Starck2002a} make use of the curvelet transform (CT) for Hubble Space Telescope (HST) image restoration from noisy data. In that paper, enhanced contrast was reported on the image of Saturn. \citet{Lambert2006} apply CT to choose significant coefficients from astroseismic observations while \citet{Starck2004} use CT for detection of non-Gaussian signatures in the observations of the cosmic microwave background. Since it is believed that CT is more suitable for representing elongated features such as lines or edges, one of the goals of this paper is to introduce CT into the Bayesian framework and show how it performs on images of planetary objects. 

The aforementioned methods work with static PSFs, i.e., they do not update the PSF of an optical system which is supplied by the operator. Nevertheless, in ground-based imaging, whether with adaptive optics (AO) or without, there are always differences between science and calibration PSFs \citep{Esslinger1998}. These differences may result from changes in seeing, wind speed, slowly varying aberrations due to gravity or thermal effects and, in AO imaging, from differential response of the wavefront sensor to fluxes received from the science and calibration objects. Quality of AO images is quantified with the Strehl ratio (SR) which is the ratio of the measured peak value of a point-source image to that of the diffraction-limited PSF, often given in percent. A perfect diffraction-limited image has SR of 100\% while a seeing-degraded image on a large telescope can have SR lower than 1\%. In this paper we will use SR to quantify the mismatch between the target and calibration PSFs. This commonly-occurring mismatch has prompted optical scientists, especially those working on AO systems, to investigate blind and myopic image restoration schemes \citep[e.g.][]{Lane1992,Thiebaut1995}. A blind method works without any information about the PSF while a myopic approach relies on some initial PSF estimate which is then updated until a solution for both the object and the PSF is found.

\citet{Conan1999} compared their myopic approach to a basic, \textquotedblleft unsupervised\textquotedblright$\,$Richardson-Lucy scheme and found, not surprisingly, that the myopic deconvolution is more stable. Nevertheless, to our knowledge there have not been any thorough and fully-fledged efforts to compare the performance of modern, static-PSF approaches to blind/myopic methods in the context of AO imaging, and so the preference for the latter algorithms still has to be justified. Our goal is to partially fill this gap. At this stage we want to mention that this article is a companion paper to \citet{BaenaGalle2011} where we showed that a modern, static-PSF code is capable of extracting accurate differential photometry from AO images of binary stars even when given a mismatched PSF. In the current paper we extend the analysis to a more complex object and we analyze the effect of noise as well as that of the mismatched PSF. We show the importance of noise control, specifically how advanced noise suppression can set off the lack of PSF-update capability in the case of very noisy observations. More generally, we want to hint at the opportunities for exchange of ideas between the communities preferring myopic and static-PSF approaches. 

The paper is organized as follows: in Section 2, the algorithms AWMLE, ACMLE, MISTRAL and IDAC are described. These algorithms represent different philosophies with regards to the deconvolution problem. AWMLE and ACMLE perform a classical static-PSF deconvolution within the wavelet or curvelet domain, MISTRAL is intended for myopic use, and IDAC can be used as a blind or a myopic algorithm. Section 3 describes the dataset we used and how we applied each of the four algorithms. {A description of the mean squared error (MSE) and the structural similarity index (SSIM) used to compare the reconstructed images can be found in section 4. Section 5 presents the comparison of two strategies of multiresolution support. The performances of the wavelet and the curvelet transforms are also discussed. Section 5 also contains the performance comparison of the four algorithms mentioned above. Section 6 summarizes and concludes the paper.}

\section{DESCRIPTION OF THE ALGORITHMS}

This section is not intended to describe the algorithms in detail but rather to offer a brief overview of their characteristics and their historical uses and performances. We would also like to point out that in the text to follow we {do not} keep the nomenclature originally used by the respective authors. {We make the following attempt at standardization: the object, or unknown, is represented with $\mathbf{o}$, the image or dataset is $\mathbf{i}$, and the PSF is $\mathbf{h}$. Upper-case notation is used for Fourier representation. The two-dimensional pixel index is $\mathit{r}$ while $\mathit{f}$ is spatial frequency. Index of a wavelet/curvelet scale is represented with $\mathit{j}$ while a single frame in a multi-frame approach is indexed with $\mathit{i}$. Finally, the symbol $\,$ $\hat{.}$ $\,$ is used for estimates. }

\subsection{AWMLE}

The Adaptive Wavelet Maximum Likelihood Estimator (AWMLE) is a Richarson-Lucy-type algorithm \citep{Richardson1972,Lucy1974} that maximizes the likelihood between the dataset and the projection of a possible solution onto the data domain, considering a combination of the Poissonian shot noise, intrinsic to the signal, and the Gaussian readout noise of the detector. This maximization is performed in the wavelet domain. The decomposition of the signal into several channels allows for various strategies to be used depending on a particular channel's scale. This is a direct consequence of the fact that in a WT decomposition the noise, together with the finest structures of the signal will be transferred into the high-frequency channels while coarse structures will be transferred into the low-frequency channels. 

The general mathematical expression that describes AWMLE is:

\begin{equation}
\mathbf{\hat{o}^{(n+1)}} = K \mathbf{\hat{o}^{(n)}} \left[ \mathbf{h^{-}} \ast \dfrac{\sum_{j} \left(  {\mathit{\omega}}_{j}^{\mathbf{{o}_h^{(n)}}} + \mathbf{m_{j}} ({\mathit{\omega}}_{j}^{\mathbf{i'}} - {\mathit{\omega}}_{j}^{\mathbf{{o}_h^{(n)}}}) \right)}{ {\mathbf{h^{+}} \ast \mathbf{\hat{o}^{(n)}}}} \right]
\label{eq10}
\end{equation}


where $\mathbf{{o}}$ is the object to be estimated, $\mathbf{h^{+}}$ is the point spread function (PSF) which projects the information from the object domain to the image domain, while $\mathbf{h^{-}}$ is its reverse version performing the inverse operation, {i.e., the conversion from the image domain to the object domain}. The so-called direct projection $\mathbf{{o}_h}$ is an image of the object projected onto the data domain by means of the PSF at iteration $\mathit{n}$, {it appears in the numerator of equation} \ref{eq10} {already decomposed in its wavelet representation.} The parameter $K=\dfrac{\mathbf{i}}{\mathbf{\hat{o}^{(n+1)}}}$ is a constant to conserve the energy.

{The variable $\mathbf{i'}$ is a modified version of the dataset which appears here because of the explicit inclusion of the readout Gaussian noise into the Richardson-Lucy scheme} \citep{Nunez1993}. {This represents a pixel-by-pixel filtering operation in which the original dataset is substituted by this modified version. It must be calculated for each new iteration by means of the following expression:}

\begin{equation}
 i'(r) = \dfrac{\sum_{k=0}^{\infty} (k e^{-(k-i(r))^2/2\sigma^2} [({{o}_h}(r))^k/k!])}{\sum_{k=0}^{\infty} ( e^{-(k-i(r))^2/2\sigma^2} [({o}_h(r))^k/k!]) }
\label{eq15}
\end{equation}
 
{Where $\mathbf{i}$ is the original dataset, $\mathit{r}$ is the pixel index, $\sigma$ is the standard deviation of the readout noise and $\mathbf{{o}_h}$ is the direct projection of the object onto the data domain. In the absence of noise ($\sigma \rightarrow 0$) the exponentials in equation} \ref{eq15} {are dominant when $k=i(r)$ and then $i'(r) \rightarrow i(r)$, i.e., the modified dataset converges to the original one. We note that by its definition, $\mathbf{i'}$ is always positive while $\mathbf{i}$ can be negative due to the presence of readout noise.}

In expression \ref{eq10}, the symbols ${\mathit{\omega}}_{j}^{\mathbf{{o}_h^{(n)}}}$ and ${\mathit{\omega}}_{j}^{\mathbf{i'}}$ correspond to wavelet coefficients, in channel $\mathit{j}$, of the multiscale representation of the direct projection $\mathbf{{o}_h^{(n)}}$ as well as that of the modified dataset $\mathbf{i'}$, respectively. 

Of particular importance is the probabilistic mask $\mathbf{m_{j}}$, centered on pixel $\mathit{r}$, which locally determines whether a significant structure is present or not in channel $\mathit{j}$. If the probability of finding a source within the mask is considered to be high then $\mathbf{m_{j}}$ will be close to 1, and so the dataset $\mathbf{i'}$ will be the only remaining term in the numerator and will be compared, at iteration $\mathit{n}$, with the estimated object at that iteration $\mathbf{{o}^{(n)}}$. If the presence of a signal within the mask is considered to be insignificant, then $\mathbf{m_{j}}$ will decrease to zero and the object will be compared with itself (by means of its projection $\mathbf{{o}_h^{(n)}}$), so the main fraction in equation \ref{eq10} will be set to unity (at channel $\mathit{j}$ in the vicinity of pixel $\mathit{r}$) and this would stop the iterative process. Therefore, the probabilistic mask $\mathbf{m_{j}}$ is used to effectively halt the iterations in a Richardson-Lucy scheme. In other words, it allows a user to choose from a wide range of the maximum number of iterations to be executed, where the reconstructions corresponding to these different numbers will not exhibit significant differences between them, which arise in the classic Richardson-Lucy algorithm due to the amplification of noise.

AWMLE in the wavelet domain was first introduced by \citet{Otazu2001}. \citet{BaenaGalle2011} presented the algorithm in the context of AO imaging and obtained differential photometry in simulated AO observations of binary systems. The accuracy of aperture photometry performed on the deconvolution residuals was compared with the accuracy of PSF-fitting, a classic approach to the problem of overlapping PSFs from point sources \citep{Diolaiti2000}. It was proven that AWMLE yields similar, and often better, photometric precision compared to StarFinder independently of the stars' separation. Even though AWMLE does not update the PSF as it performs deconvolution, it was shown that the resulting photometric precision is robust to mismatches between the science and the calibration PSF up to 6\% in terms of difference in SR which was the maximum tested mismatch. 

\subsection{ACMLE}

{Equation} \ref{eq5} {offers a direct reconstruction formula for the wavelet domain, i.e., the sum of all the wavelet planes (and the residual one) in which the original image had been decomposed allows one to obtain back that same original image. This explains the presence of summation in the numerator of expression} \ref{eq10}, {where a combination of different wavelet coefficients, some belonging to the direct projection of the object ($\mathbf{{o}_h}$) and some belonging to the modified dataset ($\mathbf{i'}$), creates the correction term in this Richardson-Lucy scheme.}

{In the curvelet domain, the reconstruction formula does not have this simple expression. For both: the forward and the inverse transforms, double Fourier inversions and complex operations with arrays are required} \citep{Candes2006}. {Nevertheless, there is an intuitive way to obtain a kind of \textit{\`a trous curvelet transform}, a method to recover the original image by using a simple sum as it is done in equation} \ref{eq5}. {This would consist of decomposing the image into $N$ curvelet scales, setting to zero all the coefficients except those corresponding to a certain scale, and applying the inverse curvelet transform to transfer the remaining coefficients to the spatial domain. One by one, the same operation would be targeted at the other of curvelet scales. }

{This would create $N$ images in the spatial domain, each of them representative of a certain curvelet scale. All of them could be added in order to recover the original image and, consequently, expression} \ref{eq10} {would still be valid. Unfortunately, this relatively straightforward and intuitive way to introduce CT into AWMLE equation invalidates most of the advantages of directly working in the curvelet domain. Firstly, it requires one forward transform plus $N$ inverse transforms at each iteration. Secondly, the sparsity concept, which allows one to work with only a few non-zero coefficients and is the basis of many of the multiresolution transforms, is abandoned in this naive approach. As mentioned by} \citet{BaenaGalle2010}, {CT tends to separate the smallest signal structures from the noise in the spatial domain better than WT. However, it is impossible to separate the signal from noise completely, and so the SNR of the finest curvelet plane is lower than the SNR of the finest wavelet plane, since less signal is represented in such high-frequency plane in the spatial domain (SNR here refers to the ratio of intensity between the remnants of objects in a particular plane and the finest structures created by noise during the classification process). In other words, the identification of the correct information is faster and more successful in the curvelet domain, especially in the highest frequency scales, when compared to the use of the representation of each curvelet scale in the spatial domain.}

{Given the above reservations about the usage of CT in the spatial domain we decided to work with the following expression:}

\begin{equation}
\mathbf{\hat{o}^{(n+1)}} = K \mathbf{\hat{o}^{(n)}} \left[ \mathbf{h^{-}} \ast \dfrac{\mathbb{C}^{-1} \left( \mathbb{C}({\mathbf{{o}_h^{(n)}}}) + \mathbf{m_{j}} \left(\mathbb{C}({\mathbf{i'}}) - \mathbb{C}({\mathbf{{o}_h^{(n)}}})\right) \right)} { {\mathbf{h^{+}} \ast \mathbf{\hat{o}^{(n)}}}} \right]
\label{eq17}
\end{equation}

{where $\mathbb{C}$ and $\mathbb{C}^{-1}$ mean, respectively, the forward and the inverse curvelet transforms. The mask $\mathbf{m_j}$ is now calculated in the curvelet domain and combines coefficients from $\mathbf{{o}_h}$ and $\mathbf{i'}$ in order to create a new curvelet correction term which is inversely transformed to the spatial domain and compared with the object estimate at each iteration. The curvelet transform used in equation} \ref{eq17} {corresponds to the so-called second generation CT and is implemented in the software CurveLab\footnote{http://www.curvelet.org}. This particular implementation of the CT exhibits a robust structure based on a mother curvelet function of only three parameters (scale, location and orientation), which is faster and simpler to use in comparison with the first generation CT based on a complex seven-index structure, which relies on the combined usage of the Starlet and the ridgelet transforms} \citep{Starck2010}.

\subsection{MISTRAL}

The Myopic Iterative STep-preserving Restoration ALgorithm (MISTRAL) is a deconvolution method within the Bayesian framework that jointly estimates the PSF and the object using some prior information about both these unknowns \citep{Mugnier2004}. This joint Maximum A Posteriori (MAP) estimator is based on the following expression:

\begin{equation}
\begin{split}
[\mathbf{\hat{o},\hat{h}}]=\mbox{arg max}[p(\mathbf{i|{o},{h}}) \times p(\mathbf{{o}}) \times p(\mathbf{{h}})] = \\ = \mbox{arg min}[J_i(\mathbf{{o},{h}}) + J_o(\mathbf{{o}}) + J_h(\mathbf{{h}})]
\label{eq20}
\end{split}
\end{equation}

where $J_i(\mathbf{o,h})=-lnp(\mathbf{i|o,h})$ is the joint negative log-likelihood that expresses fidelity of the model to the data ($\mathbf{i}$), $J_o(\mathbf{o})=-lnp(\mathbf{o})$ is the regularization term, which introduces some prior knowledge about the object ($\mathbf{o}$) and $J_h(\mathbf{h})=-lnp(\mathbf{h})$ accounts for some partial knowledge about the PSF ($\mathbf{h}$). The symbol $\mathit{p}$ in the above expressions corresponds to the probability density function of a particular variable.

MISTRAL does not use separate models for the Poisson and the readout components of noise. Instead, a nonstationary Gaussian model for the noise is adopted. What this means is that a least-squares optimization with locally-varying noise variance is employed:

\begin{equation}
 J_i(\mathbf{o,h}) = \sum_{r} \dfrac{1}{2\sigma^2(r)}[{i}(r) - ({o} \ast {h})(r) ]^2
\label{eq25}
\end{equation}

where $\mathit{r}$ stands for pixel index. This prior makes it easier to compute the solution with gradient-based techniques as compared to the Poissonian likelihood which contains a logarithm. The Gaussian assumption is typical \citep{Andrews1977} and it can be considered a very good approximation for bright regions of the image. The assumption can cause problems for low-light-level data recorded with modern CCDs of almost negligible readout noise.

The prior probability, $J_o(\mathbf{o})$, is modelled to account for objects which are a mix of sharp edges and smooth areas such as those that we deal with in this article. The adopted expression contains an edge-preserving prior that is quadratic for small gradients and linear for large ones. The quadratic part ensures a good smoothing of the small gradients (i.e., of noise), and the linear behaviour cancels the penalization of large gradients (i.e., of edges). Such combined priors are commonly called $\mathit{L}_{2}-\mathit{L}_{1}$ \citep{Green1990,Bouman1993}. The $\mathit{L}_{2}-\mathit{L}_{1}$ prior adopted in MISTRAL has the following expression:

\begin{equation}
J_o(\mathbf{o}) = \mu \delta^2 \sum_{r} \phi(\nabla {o}(r)/\delta)
\label{eq30}
\end{equation}

where $\phi(x)=|x|-ln(1+|x|)$ and where $\nabla {o}(r)=[\nabla_{x}{o}^2(r) + \nabla_{y}{o}^2(r)]^{1/2}$. Here, $\nabla_{x}\mathbf{o}$ and $\nabla_{y}\mathbf{o}$ are the object finite-difference gradients along \textit{x} and \textit{y}, respectively. Equation \ref{eq30} is effectively $\mathit{L}_{2}-\mathit{L}_{1}$ since it adopts the form $\phi(x)\approx x^2/2$ when $x$ is close to 0 and $\phi(x)/|x| \rightarrow 1$ when \textit{x} tends to infinity. The global factor $\mu$ and the threshold $\delta$ are two hyperparameters that must be adjusted by hand according to the level of noise and the object structure. Some strides towards semi-automatic setting of these parameters were made by \citet{Blanco2011}.

The regularization term for the PSF, which introduces the myopic criterion into equation \ref{eq20}, assumes that the PSF is a multidimensional Gaussian random variable. This assumption is justified when one deals with long exposures which are, by definition, sums of large numbers of short-exposures. As such they are Gaussian by the Central Limit Theorem. Adopting these conditions, $J_h(\mathbf{h})$ has the form:

\begin{equation}
J_h(\mathbf{h}) = \dfrac{1}{2} \sum_{f} \dfrac {|{{H}}(f) - {{H}_{m}}(f)|^2}{E[|{{H}}(f) - {{H}_{m}}(f)|]^2}
\label{eq40}
\end{equation}

This prior is expressed in the Fourier domain whereby upper-case notation denotes Fourier transformation. The term ${{H}_{m}}(f) = E[{{H}}]$ is the mean transfer function and $E[|{{H}}(f)-{{H}_{m}}(f)|]^2$ is the associated power spectral density (PSD) with $f$ denoting spatial frequency. This Fourier-domain prior bears some resemblance to equation \ref{eq25}. Indeed, is also assumes Gaussian statistics and it draws the solution, in the least-squares fashion, towards the user-supplied mean PSF while obeying the error bars given by the PSD (which give the variance at each frequency). The PSF prior leads to band-limitedness of the PSF estimate because the ensemble average in the denominator, $E[|{{H}}(f)-{{H}_{m}}(f)|]^2$, should be zero above the diffraction-imposed cut-off.

In practice equation \ref{eq40} relies on the availability of several PSF measurements. The mean PSF and its PSD are estimated by replacing the expected values ($E[ . ]$) by an average computed on a PSF sample. When such a sample of several PSFs is not available, as we have assumed in our work, then $\mathbf{{H}_{m}}$ is made equal to the Fourier transform of the single supplied PSF, and $E[|{{H}}(f)-{{H}_{m}}(f)|]^2$ is computed as the circular mean of $|\mathbf{{H}_{m}|^2}$. These relatively large error bars are intentional: they account for the lack of knowledge about the PSD when given only a single PSF measurement. 

In the original MISTRAL paper \citep{Mugnier2004} the code was presented mainly in the context of planetary images, for which the object prior (equation \ref{eq30}) was developed. The experimental data presented therein was obtained on several AO systems and covered a wide range of celestial objects such as Jupiter’s satellites Io and Ganymede and the planets Neptune and Uranus. MISTRAL was applied to the study of the asteroids Vesta \citep{Zellner2005} and 216-Kleopatra \citep{Hestroffer2002}, and was also used to monitor surface variations on Pluto over a 20-year period \citep{Storrs2010}.

\subsection{IDAC}

Multi-Frame Blind Deconvolution (MFBD) \citep{Schulz1993,Jefferies1993} is an image reconstruction method relying on the availability of several images of an object. In addition, many of the MFBD algorithms rely on short exposures. This was originally dictated by the notion that in imaging through turbulence short-exposure images contain diffraction-limited information while the long exposures do not \citep{Labeyrie1970}. Before the advent of AO the only way to obtain diffraction-limited data from the ground was to record short exposures which could then be processed by one of several \textquotedblleft speckle imaging\textquotedblright$\,$ methods \citep{Knox1974,Lohmann1983}. Therefore, MFBD was originally proposed in the context of speckle imaging. 

The MFBD code we use in this paper is called IDAC, for Iterative Deconvolution Algorithm in C \footnote{http://cfao.ucolick.org/software/idac/}, and it is an extension of the Iterative Blind Deconvolution (IBD) algorithm proposed by \citet{Lane1992}. IDAC performs deconvolution by numerically minimizing a functional which is composed of four constraints:

\begin{equation}
\epsilon = E_{im} + E_{conv} + E_{bl} + E_{Fm}
\label{eq50}
\end{equation}

where 

\begin{equation}
 E_{im} = \sum_{r\in \gamma}[\hat{o}(r)]^2 + \sum_{i=1}^{M}\sum_{r\in \gamma} [\hat{h}_{i}(r)]^2
\label{eq60}
\end{equation}

is the image domain error which penalizes the presence of negative pixels ($\gamma$) in both the object ($\mathbf{o}$) and the PSF ($\mathbf{h}$) estimates. The subscript $\mathit{i}$ refers to an individual data frame.

The so-called convolution error is:

\begin{equation}
 E_{conv} = \dfrac{1}{N^2} \sum_{i=1}^{M}\sum_{f} |I_i(f) - \hat{O}(f)\hat{H}_i(f)|^2 B_i(f)
\label{eq70}
\end{equation}

which quantifies the fidelity of the reconstruction ($\mathbf{\hat{O}}$) to the data ($\mathbf{I}$) in the Fourier domain. The term $B_i$ is a binary mask that penalizes frequencies beyond the difraction-imposed cut-off, $N^2$ is a normalization constant where $N$ is the number of pixels in the image.

The third constraint is called the PSF band-limit error and it is defined as:

\begin{equation}
 E_{bl} = \dfrac{1}{N^2} \sum_{i=1}^{M}\sum_{f} |\hat{H}_i(f)|^2 B'_i(f)
\label{eq72}
\end{equation}

It prevents the PSF estimates from converging to a $\delta$ function and the object estimate from converging to the observed data. The term $B'_i$ is a binary mask that is unity for spatial frequencies greater than $1.39$ times the cut-off frequency and zero elsewhere. 

The last constraint is the Fourier modulus error:

\begin{equation}
 E_{Fm} = \dfrac{1}{N^2} \sum_{f} [|\hat{O}(f)|-|{O}_e(f)|]^2 \Phi(f)
\label{eq75}
\end{equation}

where $O_e$ is a first estimate of the Fourier modulus of the object obtained, e.g., from Labeyrie's speckle interferometry method \citep{Labeyrie1970} and $\Phi$ is a signal-to-noise filter. \citet{Jefferies1993} show with several simulations that this constraint is especially important since it incorporates relatively high SNR information from a complete dataset formed by many frames. On the other hand, Fourier modulus of an object is only recoverable directly if one has a set of speckle images of an unresolved source to be used as a reference. There are workarounds to this problem, most notably reference-less approaches due to \citet{Worden1977} and \citet{vonderLuhe1984}, but these solutions are applicable only to non-compensated imaging and could not be used in our tests.

While MFBD algorithms were initially proposed in the context of speckle imaging, there is nothing preventing their application to long-exposure AO images which now contain diffraction-limited frequencies. In principle there are some inherent advantages of working with $\mathit{M}$ frames instead of using a single, co-added long exposure. The multi-frame approach reduces the ratio of unknowns to measurements from ${2:1}$ in single-image blind deconvolution to ${M+1:M}$ in multi-frame deconvolution. On top of the PSF band-limit constraint (equation \ref{eq72}) concurrent processing of many frames means that the PSF cannot converge to the $\delta$ function (this would have yielded an object equal to the data, but the data is generally temporally variable while the object is assumed constant in MFBD, which is a good assumption in the context of astronomical imaging on short time-scales). MFBD algorithms are very successful in the case of strongly varying PSFs so that the target is easily distinguished from the PSFs. On the other hand, the goal of AO is to stabilize the PSF. This implies less PSF diversity from one observation to another so that other constraints become more useful.

The code IDAC can be regarded as a precursor but also a representative of a wider class of MFBD algorithms. We will mention here the PCID code \citep{Matson2009} which has the capability to estimate the PSFs either pixel by pixel in the image domain or in terms of a Zernike-based expansion of the phase in the pupil of the telescope. It has been shown that such a
PSF re-parameterization leads to object estimates which are less noisy and have higher spatial resolution \citep{Matson2007}.

IDAC was applied to speckle observations of the binary system Gliese 914 and in the process the secondary component was resolved into two stars \citep{Jefferies1993}. Additionally, IDAC was one of the five codes used in our study of photometric accuracy of image reconstruction algorithms \citep{Gladysz2010}.

\section{DATASET DESCRIPTION AND METHODOLOGY}

\subsection{Dataset description}\label{DatasetDescription}

\begin{figure*}
   \centering 
   \includegraphics[width=15cm]{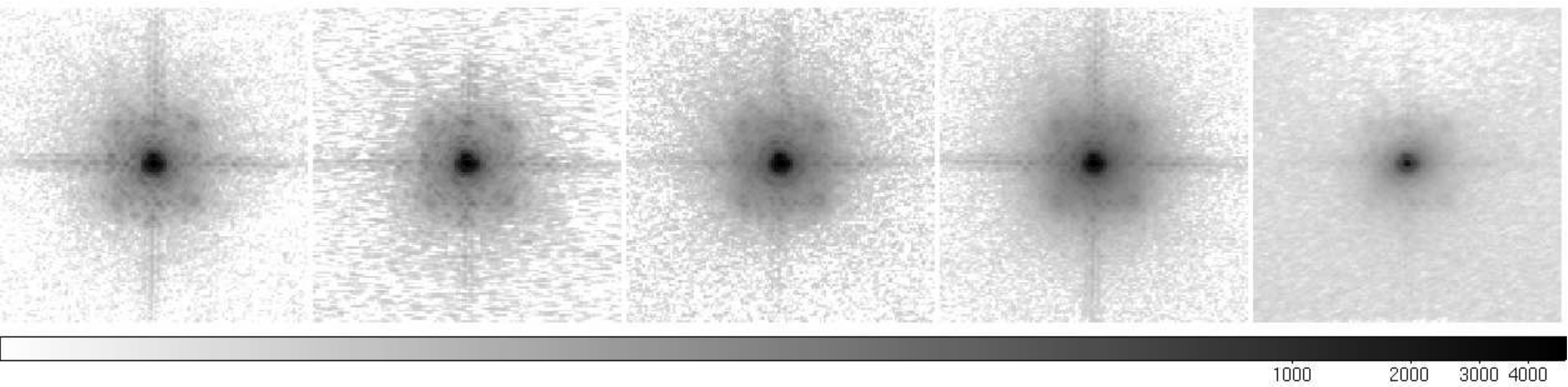}
    \caption{\footnotesize Far left: science PSF with the Strehl ratio (SR) = 53\%. Middle left: science PSF with SR = 51\%. Center: reference PSF with SR = 53\%. Middle right: reference PSF with SR = 45\%. Far right: reference PSF with SR = 36\%. Left and middle left PSFs were used to blurr the HST image of Saturn. The other three PSFs were used as reference PSFs for the algorithms. Logarithmic scale, inverted colors.}
  \label{fig15}
\end{figure*}

The image we used to test the algorithms corresponds to an observation performed with the fourth detector of the WFPC2 camera \citep{Trauger1994} (the so-called planetary camera -PC-, with a pixel size of $0.046''$) installed on the HST. This is a picture of Saturn (the whole planet is visible) with a dynamic range of up to 975 counts. In order to obtain well-defined edges and transitions between the object and the background, the image of Saturn was preprocessed so that all the pixels below a certain threshold were set to zero, thus enhancing the visibility of the Cassini and Encke divisions. This image was considered a representation of the true object.

\begin{figure*}
   \centering 
   \includegraphics[width=15cm]{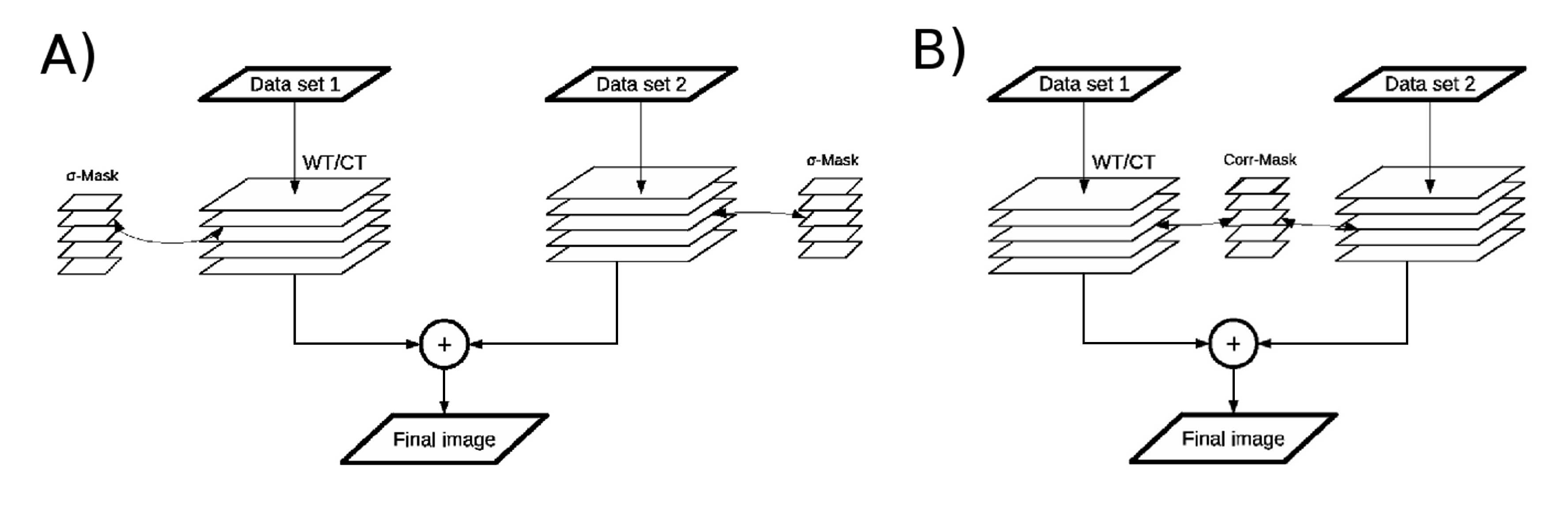}
    \caption{\footnotesize A: A flow graph illustrating the execution of the AWMLE algorithm with the $\sigma$-based mask. Datasets are decomposed by WT and each of the resulting planes is \textit{independently} analyzed with the $\sigma$-based mask. B: A respective graph for execution with the correlation-based mask. After decomposition the planes are \textit{simultaneously} analyzed with the correlation-based mask.}
  \label{fig20}
\end{figure*}

The \textquotedblleft ground truth\textquotedblright$\,$image was subsequently convolved with an AO PSF. For tests of AWMLE and ACMLE with the local-correlation mask (see Section 3.2) it was mandatory to have two blurred images (we chose PSFs with Strehl ratios equal to 51 and 53\%, see Table \ref{table:10}). This is because in these approaches two images are deconvolved simultaneously (Fig. \ref{fig20}.B). 

While the $\sigma$-based approach (see Section 3.2) does not in principle require more than one image the same two images were used again for consistency. For the same reason MISTRAL was applied to two images, instead of one, and the two separate reconstructions were averaged. In the case of MFBD restorations with IDAC we used only the 51\%-Strehl-ratio PSFs to blur the image(s). We want to remark at this stage that this requirement to have more than one image for AWMLE and ACMLE with the local-correlation mask does not fundamentally limit their applicability for astronomical imagery. A double image of the same object could be achieved in a real situation in several ways: two different observations on two consecutive nights, two outputs of a camera equipped with a beam splitter, by dividing a number of short-exposure frames into two different datasets or via the thinning method \citep{Llacer1993}, which allows an image to be split into two halves each of which preserves the Poissonian and Gaussian statistical nature of the original. 

 

The PSFs were obtained with the 1024x1024 PHARO infrared camera \citep{Hayward2001} on the 5-m Hale telescope at the Palomar Observatory \citep{Troy2000}. Closed-loop images of single stars were recorded using the $0.040'' pixel^{-1}$ mode. The images were cropped to size 150$\times$150 pixels which corresponds to a field of view (FOV) of $6''$. The observations were acquired in the K band ($2.2 \mathit{\mu}$m) where the diffraction limit is $0.086''$ so that the data meet Nyquist-sampling requirements. The filter of the observations was Brackett Gamma (BrG) and each of the PSFs in Table \ref{table:10} corresponds to a sum of 200 frames, each with an exposure time being either 1416 or 2832 ms. The individual frames were registered via iterative Fourier shifting to produce shift-and-add images \citep{BaenaGalle2011}.

\begin{table}
 
 \caption{PSFs used for the simulated observations}
 \label{table:10}
 \centering
 \begin{tabular}{c c c c}
  \hline\hline
Star & SR & exp. time (ms) & \\
  \hline
PSF1 & 53\% & 1416 & Science PSF \\
--- &  51\% & --- & Science PSF \\
  \hline
PSF2 &  53\% & 2832 & Reference PSF \\
---  &  45\%  & ---  & Reference PSF \\
PSF3 &  36\%  & 1416 & Reference PSF \\
  \hline
 \end{tabular}
\end{table}

The angular size on the sky over which the AO PSF can be assumed to be almost spatially invariant is the so-called isoplanatic angle. This parameter becomes larger at longer wavelengths. As specified by \citet{Hayward2001} the isoplanatic angle is approximately $50''$ in K-band at Palomar. The angular size of the Saturn image is $20.5''$ so we can assume the PSF remains constant throughout the FOV. The difference in pixel scales between the Saturn image and the PSFs from Palomar is very small ($0.046''$ vs. $0.040''$) and therefore we did not re-bin the images to match their pixel scales. The paper is devoted to comparison of image restoration algorithm and not to the performance evaluation of the AO system at Palomar.

We used IDAC in the multi-frame mode. One of the goals of our work was to check the trade-off between the diversity provided by more frames vs. SNR per frame. One can think of this as an exposure time optimization. For a constant total observation time per object one can use shorter exposures and hope to exploit PSF variability in subsequent image restoration with MFBD, or opt instead to use a smaller number of images with better SNR per frame. Therefore, out of the original 200-frame dataset we have produced datasets of 10, 20, 50 and 100 binned PSFs. These binned PSFs, together with the original 200-frame dataset, were used as blurring kernels for the Saturn image. For AWMLE, ACMLE and MISTRAL, which are single-image restoration codes, we only used the summed PSF. All PSFs were normalized to have total power equal to unity before the convolution procedure. 

The blurred observations were subsequently corrupted with noise. We explored three levels of noise: only readout noise of standard deviation $\sigma=1$ or $\sigma=10$, and shot noise plus readout noise of level $\sigma=10$. These levels correspond to noise which was added to the summed images (two images for AWMLE, ACMLE and MISTRAL). For MFBD with IDAC we worked with several images and decided to add to each frame the amount of noise which would result in the same SNR per summed image had the images been summed and \textit{then} processed as in the case of the other three algorithms. This means adding noise with levels: $1\cdot \sqrt{50}$, $10\cdot \sqrt{50}$ and Poisson noise plus $10\cdot \sqrt{50}$ to the 50-frame dataset, for example. This way we wanted to test the benefits of exploiting PSF diversity vs. higher noise per frame, as mentioned earlier.

One of the goals of the project was to test the susceptibility of the algorithms to mismatch between the science and calibration PSFs. Therefore, we used a matched PSF (SR=53\%), a mismatched PSF (SR=45\%) and a highly-mismatched PSF (SR=36\%) as inputs to AWMLE, ACMLE, MISTRAL and MFBD. The last SR value corresponds to half of the maximum SR in K band (63\%) predicted in simulations for the AO system in Palomar \citep{Hayward2001}. Table \ref{table:10} lists the PSFs used in the tests. Differences in SR for the same star arise because of the changing seeing. All five PSFs used in the simulations are presented in Figure \ref{fig15}.

\subsection{Methodology}

The four algorithms are very different and require different strategies to obtain the final result. For AWMLE and ACMLE, it is convenient (although not mandatory) to perform an initial decomposition of the dataset, either in the wavelet domain or in the curvelet domain. This can give the user an idea as to the number of planes to use during the deconvolution process. Additionally, such a preliminary decomposition could inform the user whether it is sensible to perform deconvolution in the highest-frequency plane where the finest structures together with noise have been classified. When working in such planes there is a trade-off between the benefit of recovering information from the finest wavelet/curvelet plane and the undesirable reconstruction of non-significant structures. Both transforms, WT and CT can also be combined into a dictionary of coefficients \citep{Fadili2006}, although for the sake of simplicity we execute them independently. 

The main difficulty when using a Richardson-Lucy-type algorithm is determining at which iteration to stop the deconvolution process. In AWMLE and ACMLE, this problem is solved by probabilistic masks. The mask can apply significance thresholds to a given location in a given plane to selectively deconvolve statistically similar regions. This concept is called multiresolution support \citep{Starck2002b}. Probabilistic masks are used to stop the deconvolution process automatically in parts of the image where significant structure cannot be discerned. We use two different kinds of probabilistic mask. The first is based on the local standard deviation within the mask itself:
\newline

 \[ m_{\sigma} =    \left\{ \begin{array}{l l}
                      	1 -exp \left\lbrace  \dfrac{-   \frac{3}{2}(\sigma_i-\sigma_{\chi})^2 }{2 \sigma_{\chi}^2} \right\rbrace & \quad if \hspace{0.15 cm} \sigma_i-\sigma_\chi > 0 \\
			0 & \quad if \hspace{0.15 cm} \sigma_i-\sigma_\chi \leq 0
                     \end{array} \right.
 \]
\begin{equation}
\label{ref53}
\end{equation}

with:
\newline

     $\sigma_i = \sqrt{\dfrac{\sum_{p\in\Phi}({\chi}_{j,p})^2}{n_f}}$  
\newline

where $\sigma_i$ is the standard deviation within the window $\mathbf{\Phi}$ centered on pixel $\mathit{i}$ of the plane ${{\chi}_j}$ (pixels within that window are indexed with $\mathit{p}$), $n_f$ is the number of pixels contained in the window and ${\sigma}_{\chi}$ is the global standard deviation in the corresponding wavelet or curvelet plane. This last value can be approximated by decomposing an artificial Gaussian noise image, with standard deviation equal to that in the dataset, into wavelet or curvelet coefficients.

The second mask is based on the local correlation between two images which are deconvolved simultaneously (but independently). If we have two observations of the same object, it can be assumed that the information pertaining to that object remains constant, while structures due to noise or atmospheric turbulence introduce some scatter in the information. The correlation mask measures, at each iteration, the similarity between the same region in two different images and, if noise amplification or speckle reconstruction is detected (in the sense that the local correlation is reduced), the algorithm is stopped for that region. The expression for that mask is:
\newline

 \[ m_{c} =    \left\{ \begin{array}{l l}
                      	\left( \dfrac{\sigma_{ij}}{\sigma_i\sigma_j} \right)^{\alpha} & \quad if \hspace{0.15 cm} 0<\sigma_{ij} \leq 1 \\
			0 & \quad if -1<\sigma_{ij} \leq 0
                     \end{array} \right.
 \]
\begin{equation}
\label{ref55}
\end{equation}

where $\sigma_i$ and $\sigma_j$ are again standard deviations within windows $\mathbf{\Phi}$ centered on the pixel \textit{i} of the plane ${{\chi}_t}$ of each image (both windows have the same position so $i=j$) and $\sigma_{ij}$ is the local covariance within such a window. Parameter $\alpha$ penalizes the fact that the correlation tends to 1 very quickly. We found that a relatively high value, $\alpha=24.5$, was enough to slow this down. One can see that the same correlation mask is applied to both images, which therefore undergo the same number of effective iterations across the FOV. 

There is no need to work exclusively with one of the two masks and reject the other. The standard-deviation mask and the correlation mask can be combined in the deconvolution process. In general a good strategy is to use the standard-deviation mask within the highest-frequency wavelet/curvelet planes, where small significant structures or point-like sources can stand out above the noise, and to use the correlation mask for the other planes, where distortions in large signal structures introduced by speckles can be partially corrected. 

Figure \ref{fig20} shows the flow diagram for AWMLE when either of the two masks is used. Standard-deviation masks are applied independently to either of the images (Fig. \ref{fig20}.A) while in the correlation-based approach the same mask is applied to both images simultaneously (Fig. \ref{fig20}.B). Using the correlation mask does not in itself imply a combined deconvolution of both images. The two threads in Figure \ref{fig20}.B are executed independently but, at each iteration, they make use of the same values to classify significant coefficients locally in each plane. The threads in Figure \ref{fig20}.A are also executed independently but the probabilistic masks used for either of the images can take different values. In both cases the two outputs of the deconvolution process are then averaged in order to obtain the final reconstruction.

{CurveLab software, which is used to introduce the CT into ACMLE, can perform digital curvelet decomposition via two different implementations. The first one is the so-called USFFT digital CT and the second one is known as the digital CT via wrapping. Both differ in the way they handle the grid which is used to calculate the FFT in order to obtain the curvelet coefficients. This grid is not defined in typical Cartesian coordinates but rather it emulates a polar representation, more suited to the mathematical framework defined by} \citet{Candes2006}. {The USFFT version has the drawback of being computationally more intensive with respect to the wrapping version, since the latter makes a simpler choice of the grid to compute the curvelets} \citep{Starck2010}. {Hence, for the reason of computational efficiency, the wrapping version was used in ACMLE.}

{The user has to provide CurveLab with the values of some parameters. Apart from the number of curvelet scales, the number of orientations or angles of representation in the second scale is also required as input from the user. This parameter will automatically set the number of angles for the rest of the scales. Evidently, the higher the number of orientations the longer the algorithm will need to perform CT, the higher the overall redundancy and the higher the computational cost for calculating the probabilistic masks at each scale. We decided to set this parameter to 16 as a trade-off between having a complete representation of all the possible orientations present in the image and the total execution time. Finally, a third parameter decides if the curvelet transform is replaced by an orthogonal wavelet representation at the final scale, i.e., the highest-frequency scale. This translates to less redundancy and a faster execution of the algorithm. Since we did not see significant differences in preliminary tests when choosing either one type of representation or the other, we set this parameter to zero, i.e., the final scale was represented by means of wavelets.}

MISTRAL performs the minimization process of equation \ref{eq20} by the partial conjugate-gradient method. The code requires that the user provides values for the hyperparameters $\mu$ and $\delta$ (see equation \ref{eq30}) which balance smoothing imposed by too much regularization against noise amplification. This has to be done by trial and error although the authors of the algorithm provide some suggestions in the original MISTRAL paper, namely that $\mu$ be set to around unity and $\delta$ be set to the norm of the image gradient ($\|\nabla i\|=[\sum_{p}|\nabla i(p)|^{2}]^{1/2}$). In practice the user experiments with various values of these parameters, centered on the values suggested by \citet{Mugnier2004}, and chooses the image reconstruction which is most visually appealing. In our tests we found that $\mu=10$ and $\delta=2$ yielded the best results. We let the code run for the maximum number of iterations set to $10^6$. It usually converged after $3000$ iterations.

\begin{figure}
   \includegraphics[width=9cm]{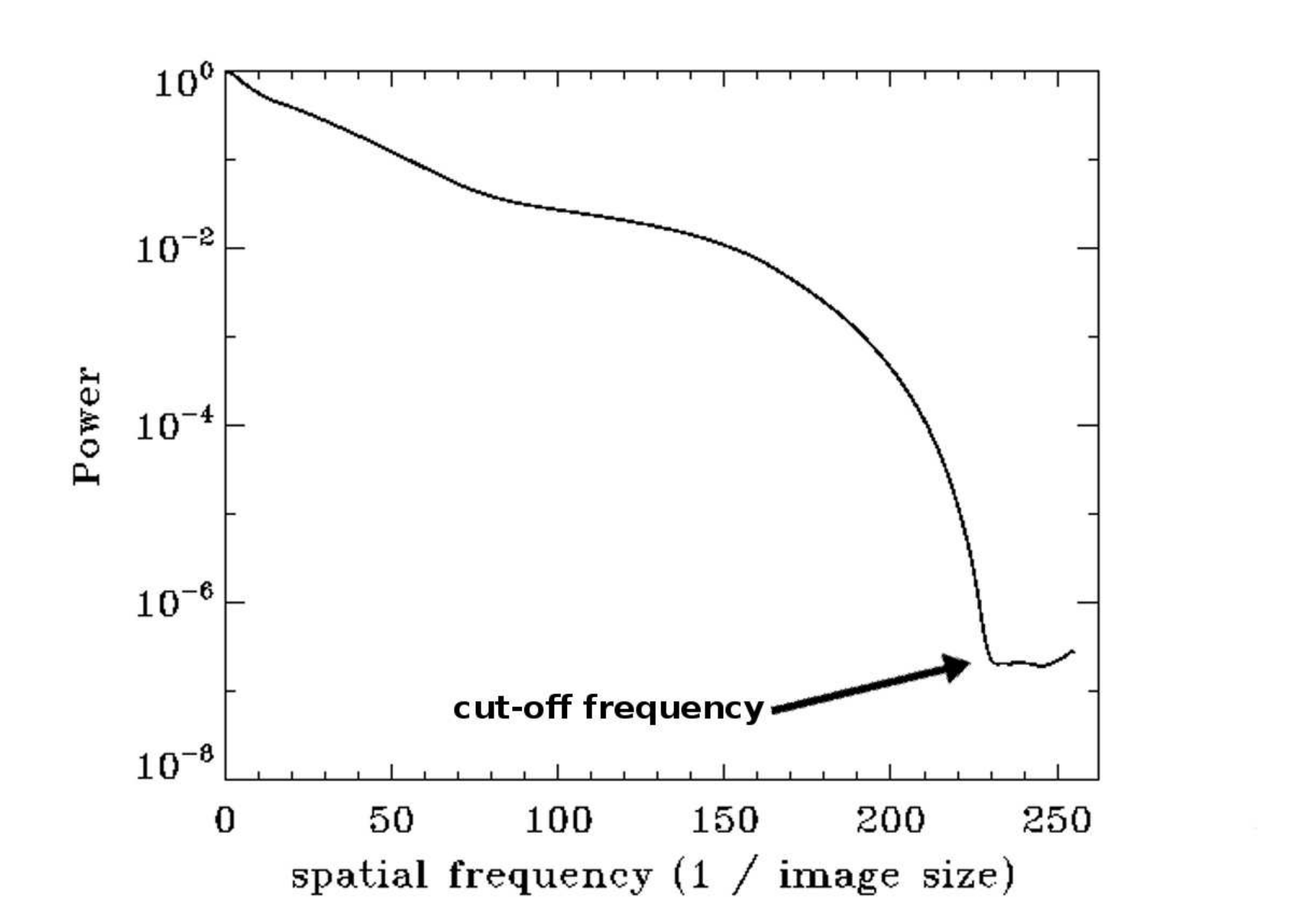}
    \caption{\footnotesize Normalized power spectrum of the 53\% SR PSF (Fig. \ref{fig15} middle left). The diffraction- and noise-limited cut-off frequency was determined to be approximately 230 pixels. The PSF was embedded in an array of zeros to match the size of the science image (512$\times$512 pixels).}
  \label{fig30}
\end{figure}

IDAC requires the user to provide the value of the diffraction-limit cut-off. Caution should be taken here: the code requires that all the supplied images be of the same dimensions. Therefore, when one has a PSF image which is smaller in size than the target image, then the PSF should be embedded in an array of zeros. Subsequently, the diffraction-limit cut-off should be estimated from the zero-padded, and not the original, Fourier-transformed PSF (see Figure \ref{fig30}). Another parameter that should in theory affect the restorations, a scalar quantifying user’s confidence in the supplied PSF, was found to have negligible effect on the outputs.

As with MISTRAL we let the code run for the maximum number of iterations set to $10^6$. For images with high level of noise ($\sigma=10$ and Poisson noise plus $\sigma=10$) it converged quickly, after 15-20 iterations, which can be considered a good behaviour as noise did not become amplified. For cases with low noise ($\sigma=1$) the algorithm converged after 50-100 iterations and produced generally sharper reconstructions.

\section{{Image quality metrics}}

One of the goals of this work was to compare the quality of reconstructions yielded by several codes. Therefore, metrics were needed for image quality assessment. {Here we have chosen the mean squared error (MSE) and the so-called structural similarity index (SSIM) }\citep{Wang2004}.{ The former is probably the most commonly used metric in image processing }\citep[e.g.][]{Mugnier2004}, {On the other hand, SSIM was developed to mimic some of the hypothesized properties of the human visual system (HVS) and has been shown to overcome some of the weaknesses of the MSE.}

\subsection{Structural Similarity Index}

The principal idea underlying the structural similarity approach is that the HVS is adapted to extract \textit{structural} information from images, and therefore, a measurement of structural similarity (or distortion) should provide a good approximation to perceptual image quality (as obtained from human observers). The metric, SSIM, is a function of two images, denoted here by $\mathbf{x}$ and $\mathbf{y}$. If $\mathbf{x}$ is the reference image, then SSIM can be regarded as a measurement of the quality of the second image $\mathbf{y}$. SSIM is defined as follows:

\begin{equation}
\mbox{SSIM}(\mathbf{x},\mathbf{y}) = \left[ \frac{(2\mu_x\mu_y + C_1)(2\sigma_{xy} + C_2)}{(\mu_x^2 + \mu_y^2 + C_1)(\sigma_x^2 + \sigma_y^2 + C_2)} \right] ^{\gamma}
\label{eq80}
\end{equation}

where $\mu_x$ and $\mu_y$ are the mean values of the images $\mathbf{x}$ and $\mathbf{y}$, respectively. Parameters $\sigma_x$ and $\sigma_y$ are the standard deviations and $\sigma_{xy}$ is the covariance between the two images. The metric is actually a multiplication of three functions that compare, respectively, the luminance, contrast and structure of the two images; i.e., $SSIM(x,y)=[l(x,y)^{\alpha_1} \cdot c(x,y)^{\alpha_2} \cdot s(x,y)^{\alpha_3}]$. These functions have the following expressions:

\begin{equation}
l(x,y) =  \frac{(2\mu_x\mu_y + C_1)}{(\mu_x^2 + \mu_y^2 + C_1)} 
\label{eq82}
\end{equation}

\begin{equation}
c(x,y) =  \frac{(2\sigma_x\sigma_y + C_2)}{(\sigma_x^2 + \sigma_y^2 + C_2)}  
\label{eq84}
\end{equation}

\begin{equation}
s(x,y) =  \frac{(\sigma_{xy} + C_2/2)}{(\sigma_x + \sigma_y + C_2/2)} 
\label{eq85}
\end{equation}

The functions can be weighted in order to adjust the relative importance of each of them. For the sake of simplicity, we always have $\alpha_1=\alpha_2=\alpha_3=1$, which yields expression \ref{eq80}. It is worth to mention that these three functions are relatively independent, which is physically logical since a change in luminance and/or contrast of the imaged scene should not impact the perception of the structure of an object.

The SSIM has the properties of: symmetry, i.e., $SSIM(x,y) = SSIM(y,x)$; boundedness, whereby $SSIM(x,y) = 1$ indicates that the two images are identical; and uniqueness, in the sense that the previous upper bound is only reached when $x=y$. Note that $SSIM(x,y)$ can take negative values if $x$ and $y$ have an inverse correlation between them, i.e., $\sigma_{xy} < C_2/2$.

Parameters $C_1$ and $C_2$ are intended to ensure the results are stable when either $(\mu_x^2 + \mu_y^2)$ or $(\sigma_x^2 + \sigma_y^2)$ are close to zero or when $(\mu_x\cdot\mu_y\cdot\sigma_{xy})$ is identically zero, as is the case for pixels belonging to the background in the original image. The inclusion of background pixels between Saturn's body and the innermost ring, or in the vicinity of the outermost ring, is motivated by the behaviours of the deconvolution algorithms in these regions, as will be discussed later. Additionally, including background pixels gives a measurement of the ability of an algorithm to remove from the background (and reintroduce in the object) the light scattered by the PSF. 

The choice of $C_1$ and $C_2$ is somehow arbitrary but, as commented by \citet{Wang2004}, the performance of the SSIM index is almost insensitive to variations of these parameters. We decided to set $C_1$ to $300$ and $C_2$ to $30$ in order to assign more weight to those pixels that belong to the planet, rather than to the background. Nevertheless, in the initial tests we have not encountered substantial differences in the appearance of SSIM maps for different values of both $C_1$ and $C_2$. 

As suggested by its proponents, it is preferable to apply SSIM locally, by means of sliding windows centered on a given pixel, rather than over the entire image, in order to visualize space-variant features or distortions \citep{Wang2006}. Furthermore, such quality maps can provide more information about the local degradation of the image. Parameter $\gamma$ in equation \ref{eq80} is used to enhance the visibility of such maps. In any case, it is also possible to compute a single overall quality measure of the entire image by averaging all values calculated at the individual local windows. This average is called the mean structural similarity index (MSSIM):

\begin{equation}
\mbox{MSSIM}(x,y) = \frac{1}{M} \sum_{j=1}^M SSIM(x_j,y_j) 
\label{eq86}
\end{equation}

Where $M$ is the number of local windows. At this point, we want to remark that the MSSIM metric must not be viewed as an \textquotedblleft absolute truth\textquotedblright$\,$ about the quality of the different reconstructions. Since Saturn image has zones with very different characteristics (e.g., background, main body, rings and planetary poles) MSSIM should be seen as a rough overview of an algorithm's performance, but an inspection of the SSIM maps and reconstructions themselves is essential as well.

\begin{figure}
   \centering 
   \includegraphics[width=9cm]{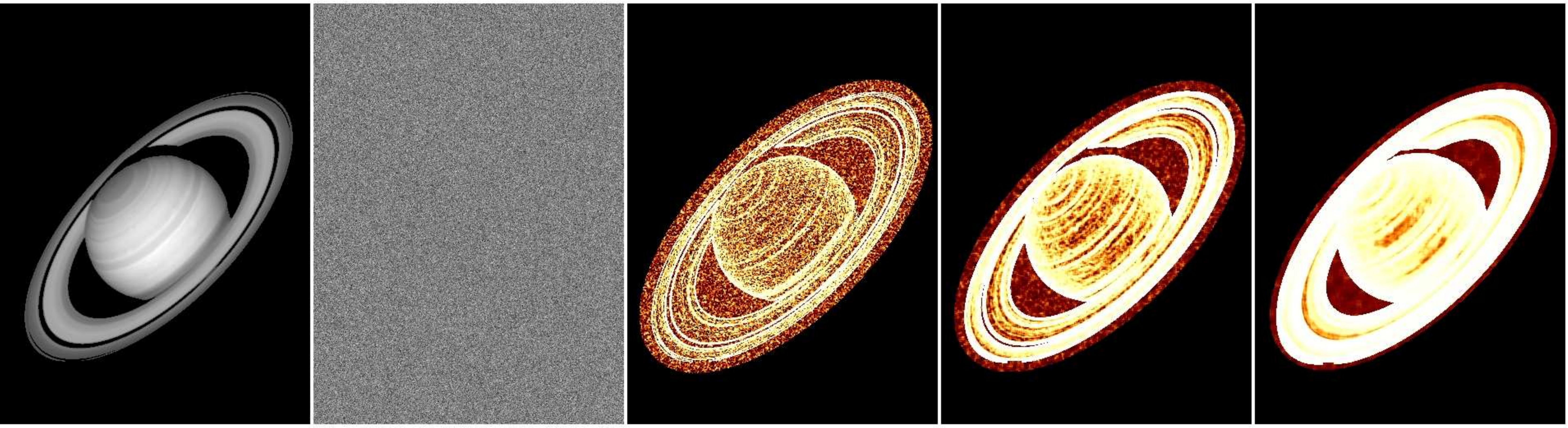}
    \caption{\footnotesize Far left: original image. Middle left: white Gaussian noise ($\sigma=10$) or absolute error map of the distorted image. Center: SSIM map of the distorted image created with 2$\times$2 sliding windows. Middle right: SSIM map of the distorted image created with 4$\times$4 sliding windows. Far right: SSIM map of the distorted image created with 8$\times$8 sliding windows. Higher brightness indicates better quality. For a given position the structural similarity index was calculated only if all the pixels in a window belonged to the object and/or to the background in the vicinity of the object.}
  \label{fig40}
\end{figure}

Figure \ref{fig40} shows the SSIM maps produced for the Saturn image after adding white Gaussian noise. The reader may note that SSIM indicates worse quality for the smooth parts of the object, for example the main body of Saturn, than for the edges. This attribute of SSIM correlates well with perceptual image quality which is known to depend on the so-called contrast masking effect \citep{Wang2006}. What this means is that the visibility of the noise-corrupted signal, as perceived by human observers, varies significantly at different locations for objects which are a combination of smooth areas, textures and sharp edges. This can be quantified using a space-variant metric. On the other hand, a global metric such as MSE, even when applied locally, would yield the same results for various parts of the image shown in Figure \ref{fig40}. Of course, this ability of SSIM to capture local image quality depends on the size of the sliding window, as can be seen by comparing the three rightmost panels of Figure \ref{fig40}. 

We note that SSIM is a widely-used metric for testing image quality in the remote sensing applications \citep{GonzAudicana2005,Otazu2005}.

\subsection{{Mean Squared Error}}

{The mean squared error is probably the oldest and most often used metric to evaluate the resemblance of two images. It is defined as:}

\begin{equation}
\mbox{MSE} = \frac{1}{M} \sum_{j=1}^M (x_j - y_j)^2
 \label{eq90}
\end{equation}

{Where, in this context, $M$ is the total number of pixels of the object and in the sorrounding area. It is also very common to use the related metric, the peak signal-to-noise ratio (PSNR), defined by:}

\begin{equation}
\mbox{PSNR} = 10 \log_{10} \dfrac{L^2}{MSE}
 \label{eq95}
\end{equation}

{Where $L$ is the dynamic range of the image. These metrics are very easy to compute and measure the pixel-by-pixel departure between the reconstruction and the reference object. The MSE and PSNR have a clear physical meaning: they quantify the energy of the error. Hence, they are well suited to the task of estimating absolute photometric error between the two images.} 

{However,} \citet{Wang2006} {show that MSE exhibits poor correlation with the real (as perceived by humans) image quality, e.g., important modifications of a reference image, like defocusing or Gaussian noise contamination, produce relatively stable MSE values, whereas small spatial shifts or rotations, while having negligible effect on the subjective image quality, can yield very different MSEs or PSNRs. In the context of the Richardson-Lucy approach, where noise is amplified after a certain number of iterations, MSE and PSNR metrics can be insensitive to such amplification. Large visual distortions produced by noise, that might invalidate the scientific benefit of the reconstruction, could appear transparent for MSE because pixel-by-pixel absolute departure from the true value has not to be necessarily very high.}

{On the other hand, SSIM measures the local resemblance between the deconvolved image and the reference image by means of variance and correlation terms. Therefore, it is very sensitive to the presence of noise and distortions produced by defocusing, which are the typical distortions present in the deconvolution problem we are dealing with in this paper. On the other hand, we have detected the weakness of SSIM in detecting luminance changes (photometric differences). Although in principle it should provide a measurement of the photometric accuracy of reconstructions by means of equation} \ref{eq82}{, in practice, relatively large departures between both images do not necessarily produce poor SSIM values if such images show the same visual structure or trend in their respective numerical ranges. }

{Summarizing, MSE and SSIM metrics can be considered complementary to each other. Therefore, usage of both metrics constitutes a sound methodology for evaluating the performance of different deconvolution algorithms. The former quantifies photometric/radiometric departures more accurately, while the latter is more sensitive to the loss of perceivable image quality which can be the result of, e.g., noise amplification or residual blur.}

{In general, we have preferred to use PSNR rather than MSE to quantify differences between images, since the former has the same tendency as SSIM, i.e., higher values correspond to better reconstructions. This common property facilitates easy side-by-side comparisons. We set $L=975$ in equation} \ref{eq95} {to match the dynamic range of the \textquotedblleft ground truth\textquotedblright$\,$ object. PSNR maps were generated using expressions} \ref{eq90} and \ref{eq95} {but on a pixel-by-pixel basis in order to show the behaviour of this metric locally}.

\section{RESULTS AND DISCUSSION}

\subsection{Correlation-based mask versus $\sigma$-based mask}\label{RyD_masks}

\begin{figure}
   \centering 
    \includegraphics[width=8cm]{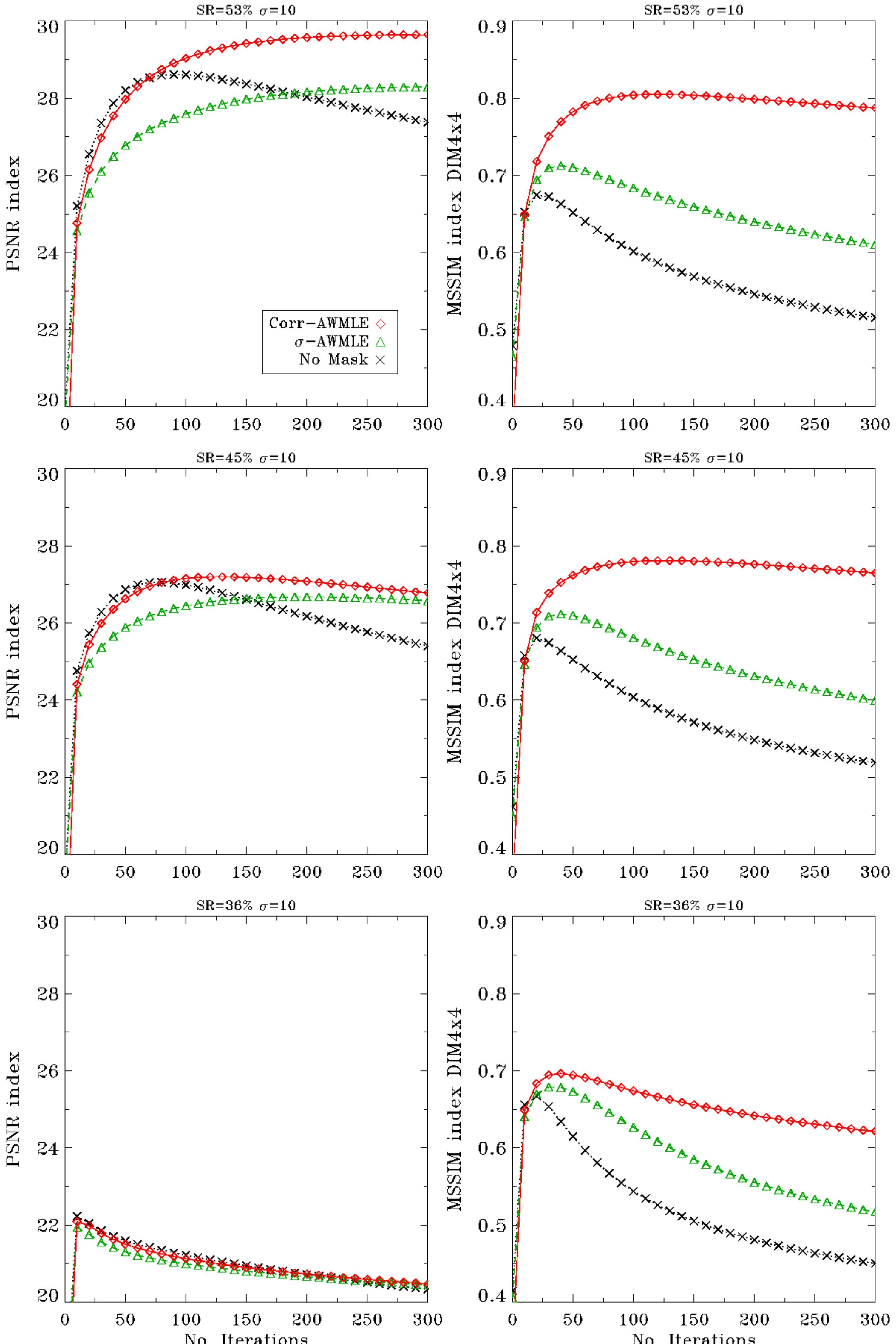}
    \caption{\footnotesize {Top row: evolution of the PSNR and MSSIM with respect to the number of iterations for the Saturn image degraded and corrupted with Gaussian noise ($\sigma=10$) and using the SR=53\% reference PSF. The left column shows PSNR index.} The right column shows MSSIM index measured within 4$\times$4 sliding windows. Middle row: SR=45\% reference PSF. Bottom row: SR=36\% reference PSF. Diamonds: AWMLE with correlation-based mask. Triangles: AWMLE with standard deviation-based mask. Crosses: AWMLE with no mask.}
  \label{fig50}
\end{figure}

The tests of the two types of probabilistic masks were conducted using AWMLE in the wavelet domain, which was the framework used by \citet{Otazu2001} to test his original design, i.e., AWMLE with $\sigma$-based mask. The algorithm was applied to two images, corresponding to the 51 and 53\% SR, and corrupted with white Gaussian noise of $\sigma=10$. Both images were decomposed into three wavelet planes plus a wavelet residual. As mentioned before, three levels of mismatch between the science and calibration PSF were analysed. The wavelet planes were also deconvolved without using any kind of mask which, in practice, results in the execution of the Richardson-Lucy approach albeit in the wavelet domain.

\begin{figure*}
   \centering 
    \includegraphics[width=17cm]{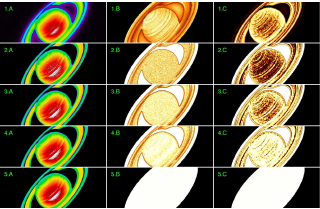}
    \caption{\footnotesize 1.A: Degraded and corrupted image with Gaussian noise ($\sigma=10$). 2.A: Reconstruction in the wavelet domain with no mask. 3.A: Reconstruction in the wavelet space with mask based on the local standard deviation. 4.A: Reconstruction in the wavelet domain with mask based on the local correlation. 5.A: Original image. {Column B: PSNR maps.} Column C: 4$\times$4 SSIM maps. All reconstructions used the SR=45\% PSF as the calibrator. The reconstructions were stopped at 100 iterations. Column A is represented on a linear scale from 0 to 975, matching the dynamic range of the original image, i.e., subpanel 5.A. {Column B is represented on a linear scale from 0 to 50.} Column C is represented on a linear scale from 0 to 1. In columns B and C higher brightness indicates better quality.}
  \label{fig60}
\end{figure*}

The algorithm was executed for 300 iterations, storing intermediate results every 10 iterations. All the outputs were analyzed by means of PSNR and SSIM, using 4$\times$4 sliding windows. SSIM was computed only in those windows where all the pixels belonged to the object or a small border around it (for the appreciation of how big this border is see Fig. \ref{fig40}). The results from such windows were averaged in order to obtain the global evolution of SSIM with respect to the number of iterations (Fig. \ref{fig50}). 

As can be seen in figure \ref{fig50}, both masks were able to control degradation in PSNR, typical to the MLE approach, across a wide range of the number of iterations. When no mask is used the general quality of reconstruction quickly drops. Although \citet{Otazu2001} assumed that the $\sigma$-based mask is able to stop noise amplification, one can observe (Fig. \ref{fig50}, right column) that this resistance is not permanent and eventually image quality can drop significantly when this mask is employed in deconvolution. {It is interesting to note that photometric accuracy of the reconstruction with the $\sigma$-based mask remains almost constant while image quality gets degraded by noise} (Fig. \ref{fig50}, row 2 and Fig. \ref{fig60}, subpanel 3.A), {thus showing the different behaviour and performance of PSNR and SSIM metrics}. The correlation mask was able to stabilize the influence of noise more effectively. Furthermore, PSNR and MSSIM were higher for the correlation mask than for the $\sigma$-based mask. For example, at iteration number 100 and when the matched 53\% SR PSF was used, {the difference in MSSIM was $\sim$15\% and the \textquotedblleft gap\textquotedblright$\,$ between both masks is a constant of $1.5$dB in PSNR. Comparing the evolutions of PSNR and MSSIM} (Fig. \ref{fig50}, first row, no mask employed) {one can see that when the reconstruction achieves its maximum photometric resemblance to the \textquotedblleft ground truth\textquotedblright, at iteration $\sim100$, amplification of the noise is already so evident that the reconstruction shows a big degradation which eventually invalidates it} (Fig. \ref{fig60}, row 2).

It should also be mentioned that a mismatch of $\sim$7\% in SR between the science PSF and the reference PSF (SR=45\%) does not lead to large differences in the results (Fig. \ref{fig50}, middle row), with respect to those obtained with a well-matched reference PSF (SR=53\%, Fig. \ref{fig50}, top row). The usage of the SR=36\% reference PSF, which implies a mismatching of $\sim$16\% SR, can be seen to have a more noticeable effect on the process of noise control (Fig. \ref{fig50}, bottom row). 

Figure \ref{fig60} shows PSNR and 4$\times$4 SSIM maps of the reconstructed images, as well as the reconstructions themselves. Results of the Richardson-Lucy method (in the sense that no mask was used, Fig. \ref{fig60}, subpanel 2.A) produced even worse SSIM maps than those obtained from the original, blurred and noise-corrupted image. This is due to visible noise amplification despite the resolution enhancement achieved from the reconstruction. The use of the masks clearly improves the results, more so if the correlation-based mask is applied in the wavelet domain (Fig. \ref{fig60}, row 4).

\begin{figure*}
   \centering 
    \includegraphics[width=17cm]{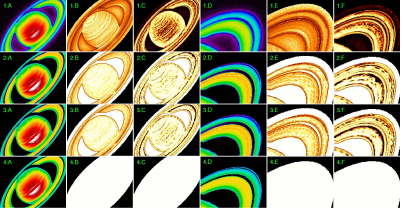}
    \caption{\footnotesize {Row 1: Degraded and corrupted image with Gaussian noise ($\sigma=10$). Row 2: Reconstruction in curvelet domain with mask based on the local standard deviation.} Row 3: Reconstruction in wavelet domain with mask based on the local correlation. Row 4: Original image. {Columns B and E: PSNR maps.} Columns C and F: 4$\times$4 SSIM maps. Column D: fragment of the image in column A. All reconstructions used the SR=45\% PSF as the calibrator. The reconstructions were stopped at 100 iterations. Columns A and D are represented on a linear scale from 0 to 975, matching the dynamic range of the original image, i.e., subpanel 4.A. {Columns B and E are represented on a linear scale from 0 to 50.} Columns C and F are represented on a linear scale from 0 to 1. In columns B, C, E and F higher brightness indicates better quality.}
  \label{fig65}
\end{figure*}

\subsection{Wavelet vs. curvelet transform}

\begin{figure}
   \centering 
    \includegraphics[width=8cm]{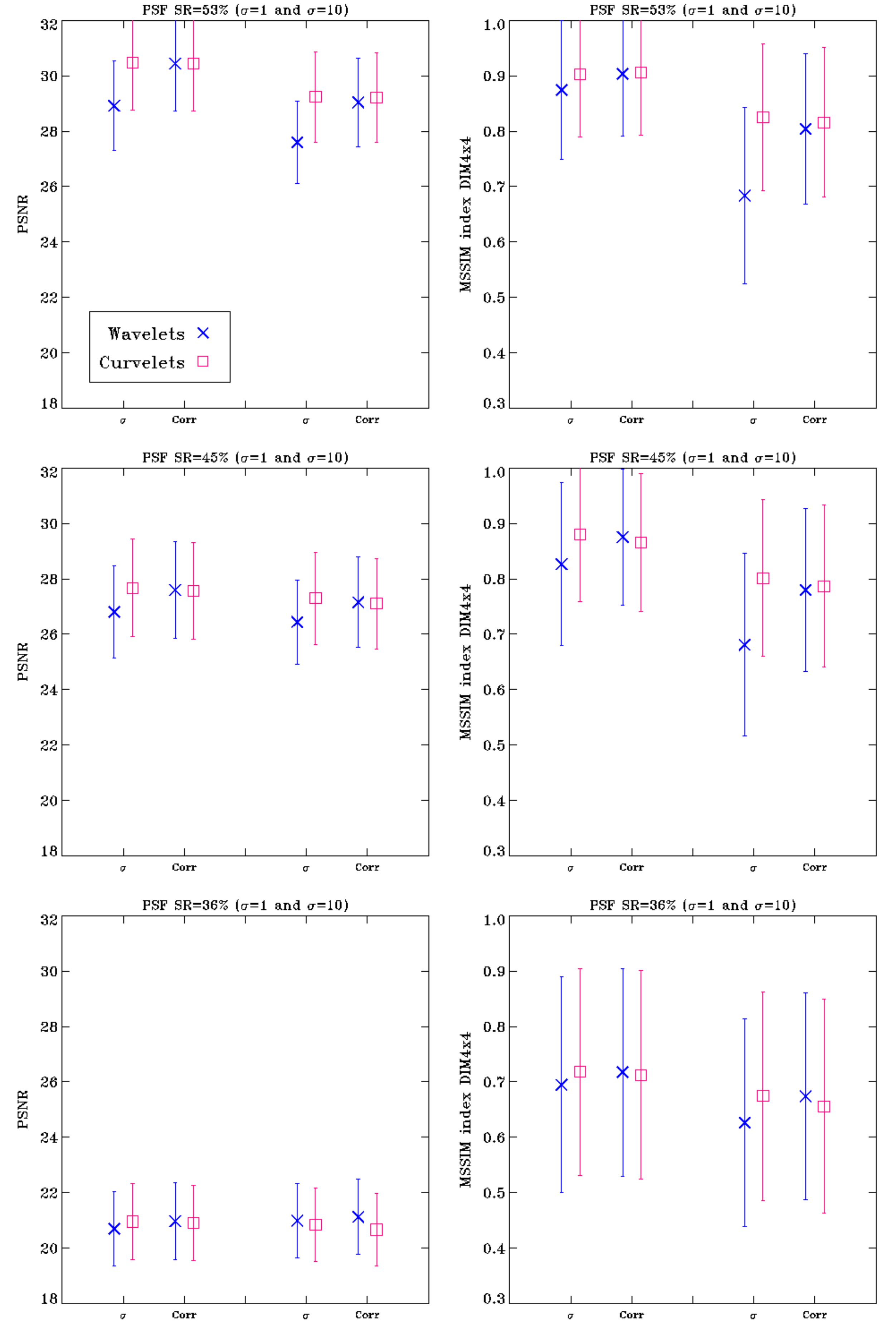}
    \caption{\footnotesize {Left: PSNR index calculated from results obtained in the wavelet and in the curvelet domain with two different types of mask (standard deviation and correlation).} Right: 4$\times$4 MSSIM index. Every plot shows results which were obtained from images contaminated with Gaussian noise of level $\sigma=1$ and $\sigma=10$. Three levels of PSF mismatch have been tested: top row SR=53\%, middle row SR=45\%, bottom row SR=36\%. Crosses: wavelet-based approach (AWMLE). Squares: curvelet-based approach (ACMLE).}
  \label{fig85}
\end{figure}

{The dataset described in Subsection} \ref{RyD_masks} {was also used to compare the quality of reconstructions when either the wavelet (WT) or the curvelet (CT) representation of the object was used. The schemes presented in equations} \ref{eq10} and \ref{eq17} {were executed in both domains. Again, the same two types of masks were employed: one based on the standard deviation and another based on the local correlation. Hundred iterations was considered to be enough to enhance the resulting image without the noise reconstruction affecting its general quality.}

{Looking at figure }\ref{fig65}, {at first glance there is no great difference between the wavelet and curvelet reconstructions. The CT and the WT produce similar PSNR and 4$\times$4 SSIM maps within the body of the planet} (Fig. \ref{fig65}, subpanels 2.B/C and 3.B/C) {whereas WT has reconstructed the inner ring slightly better (subpanels 2.E/F and 3.E/F). On the other hand, CT has brought out some elongated features present in the original image on the inner ring which are now more pronounced than in the original (subpanels 2.D and 4.D). Looking directly at the reconstructions} (Fig. \ref{fig65}, subpanels 2.A and 3.A) {one can see that CT has reconstructed lines and visual transitions more homogenously.}

{In order to determine which of the two transforms performs better as a whole, the mean value and standard deviation averaged across all the PSNR and SSIM maps (4$\times$4 windows) were computed for two noise levels ($\sigma=1$ and $\sigma=10$ readout noise). The results are presented in Figure} \ref{fig85}.

{Apart from the logical conclusions related to the general improvement corresponding to low-level noise and well-matched PSF, it is possible to observe that CT works better than WT on this dataset in most of the situations. If $\sigma$-based mask is used, CT always exhibits an improvement with respect to WT. When PSF with SR=45\% is used as calibrator, this improvement can be measured to be around $1$dB in terms of PSNR or $5$-$12$\% in terms of MSSIM depending on the noise level, thus showing that resistance to noise amplification is more robust in the curvelet domain.}

{As in section} \ref{RyD_masks}, {correlation and $\sigma$-based masks were also applied in the curvelet domain. On this occasion, we did not find large differences between the results. However, some artifacts or elongated structures were visible in the correlation mask results, especially in the presence of shot plus readout noise. Such artifacts are related with the result of wrong identifications of curvelet coefficients or non-homogenous reconstruction of different angle orientations at the same curvelet scale. This suggests that the correlation mask, in the curvelet domain, cannot be generalized for all the scales, as we do in the wavelet domain, but should be modified for each scale and each orientation.}

{The $\sigma$-based based mask in the curvelet domain produces slightly better results than the correlation-based mask in the wavelet domain. The slight decrease in reconstruction quality that can be observed between the standard-deviation and the correlation-based masks when CT is used, is related to the appearence of artifacts in the reconstructions, as was already mentioned.}

\subsection{Static-PSF approach versus blind/myopic approaches}

One of the goals of the current work was to test the static-PSF reconstruction approach (based on multiresolution support) vs. typical methods used in AO (based on the myopic/blind philosophy). Again, we used three levels of PSF mismatch as well as three noise levels: Gaussian $\sigma=1$, Gaussian $\sigma=10$ and Poissonian plus Gaussian $\sigma=10$. Therefore, the algorithms were tested in different noisy scenarios (low noise level, Gaussian dominant and Poissonian dominant). 

As mentioned in Section \ref{DatasetDescription} datasets of 10, 20, 50 and 100 binned images, together with the original 200-frame dataset, were used as input for IDAC to test the trade-off between the (implied) PSF diversity and SNR per frame. We tried to shed some light on the question which of the following strategies is better: deconvolving a long exposure image with a relatively high SNR (AWMLE, ACMLE and MISTRAL algorithms) or tackling the problem by dividing the dataset into more (diverse) frames at the expense of reducing the SNR for each frame (MFBD). 

It was found that the original 200-frame set yielded significantly worse results than the smaller sets. Specifically, the 50-frame dataset proved to be the best input to IDAC although the differences between its output and those of the 10 and 20-frame sets were small.

\begin{figure*}
   \centering
    \includegraphics[width=15cm]{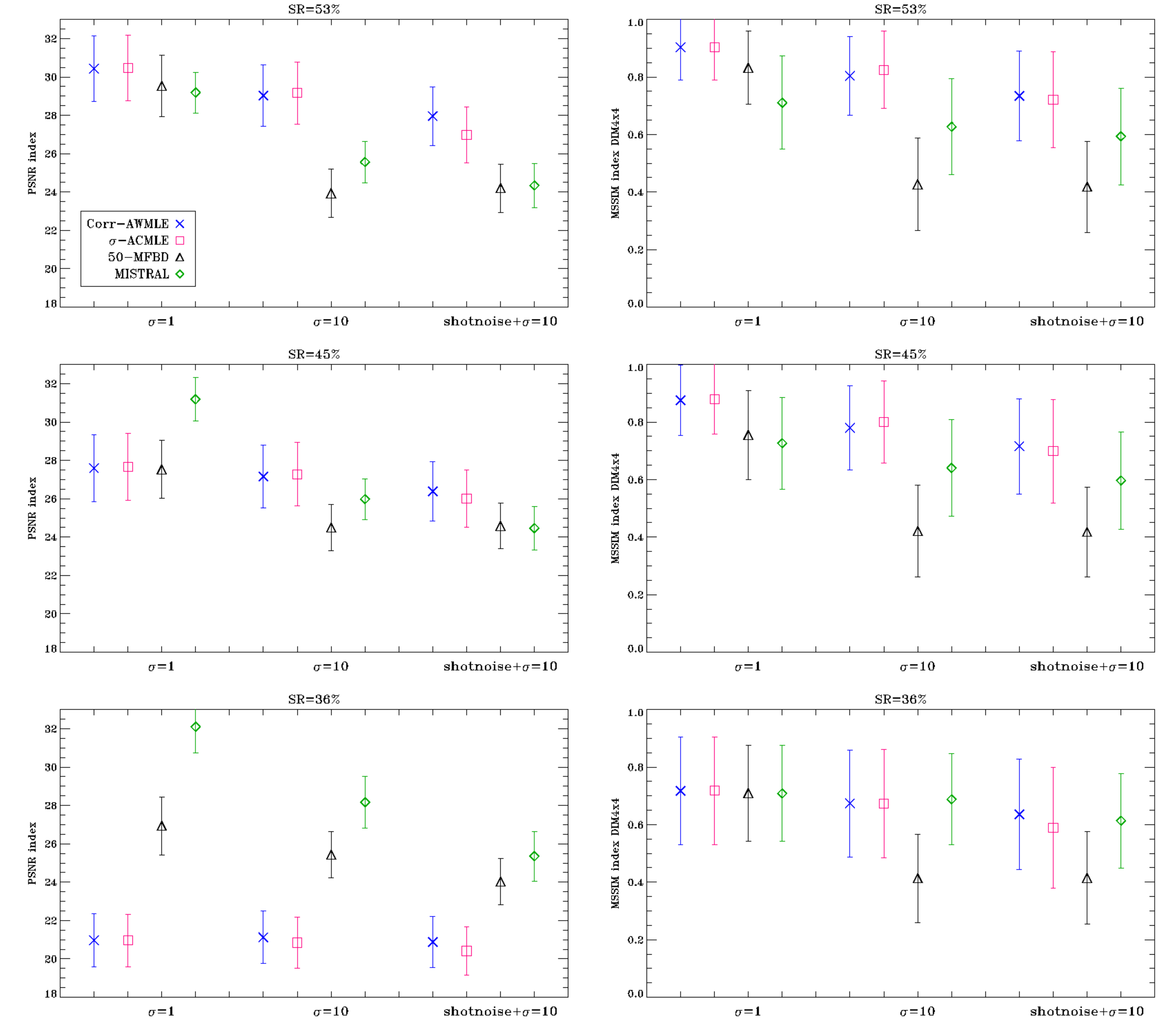}
    \caption{\footnotesize {Each plot shows the mean value and standard deviation of the PSNR and 4$\times$4 MSSIM metrics calculated from results obtained with AWMLE and ACMLE (the former using the correlation-based mask and the latter using the $\sigma$-based mask), MFBD applied over a 50-frame dataset, and MISTRAL. Deconvolution was performed with three PSFs in order to represent different levels of miscalibration. Left column: PSNR. Right column: 4$\times$4 MSSIM. Top row: dataset was deconvolved with a matched PSF (SR=53\%). Middle row: dataset was deconvolved with mismatched PSF (SR=45\%) Bottom row: dataset was deconvolved with highly-mismatched PSF (SR=36\%). Crosses: AWMLE with correlation-based mask. Squares: ACMLE with $\sigma$-based mask. Triangles: MFBD (50 frames). Diamonds: MISTRAL.}}
  \label{fig90}
\end{figure*}

\begin{figure*}
    \centering 
     \includegraphics[width=15cm]{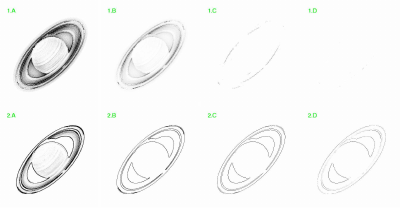}
     \caption{\footnotesize {Row 1: MSE maps of MISTRAL reconstruction. Row 2: MSE maps of ACMLE reconstruction ($\sigma$-based mask). Column A: linear scale from 0 to 2000. Column B: linear scale from 2000 to 10000. Column C: linear scale from 10000 to 20000. Column D: linear scale from 20000 to 40000. Reconstructions were obtained from datasets corrupted with Gaussian noise of $\sigma=1$. Reference PSF at SR=45\%. Inverted color scale was used to keep the general choice of representing better qualities with higher brightness. These reconstructions correspond to the particular case shown in figure }\ref{fig90}, {middle row, left column, $\sigma=1$, green diamond (MISTRAL) and pink square (ACMLE). Note that averaged PSNR for MISTRAL presents a better value due exclusively to the bad performance of the rest of algorithms in the border of the object with the zero-valued background, whereas ACMLE exhibits a better performance for the rest of pixels. } }
   \label{fig100}
 \end{figure*}

Figure \ref{fig90} shows the mean value and standard deviation of all the PSNR and 4$\times$4 SSIM windows measured between the reconstructed image and the original one. Under low-noise conditions it is possible to observe that MLE based on either wavelets or curvelets gives best results even for a PSF mismatch of 6-8\% in terms of SR. {At this stage, we want to stress the apparent contradiction between this statement and the particular case shown in figure} \ref{fig90}, {middle row, readout noise of $\sigma=1$, as well as the different behaviour between PSNR and MSSIM indices in this situation. Looking at figure} \ref{fig100}, {where MSE maps obtained from MISTRAL and ACMLE reconstructions at different scales of representation are shown, one can see that ACMLE presents better values in all the pixels of the object (planet's body, rings, space between the body and the innermost ring) except the pixels at the edges of the main planet body and the rings which show poor numbers. For these pixels, $\mathit{L}_2-\mathit{L}_1$ object prior implemented in MISTRAL is working impressively well, while ACMLE's values show large departures from the real object. In this situation, SSIM boundedness property, which fixes an upper limit of the metric, offers a better view of the algorithms performance.}

When the mismatching of the calibrator is as much as 15-17\% the performance of a static-PSF approach {deteriorates significantly} as shown in figure \ref{fig90}, bottom row, left column. The quality of reconstructions for the myopic/blind codes is more uniform with respect to the PSF mismatch which is a logical result since such algorithms are designed to deal with large differences between the science PSF and the reference. At the level of mismatch of $\sim$15\% there is already a visible qualitative difference between the PSFs (compare the leftmost and rightmost panels of Figure \ref{fig15}) and this affects the performance of static-PSF codes. {We again want to point out the contradiction between results shown by PSNR and MSSIM when PSF of SR=36\% is used. Photometric values achieved by static-PSF approaches exhibit larger departures but the visual quality is not very degraded when the reconstruction and the real object are compared (in different scales), i.e., local behaviours within each SSIM window do not show significant differences in variance and correlation. In this sense and in this case, the PSNR metric works better. These two aforementioned contradicting results from PSNR and MSSIM demonstrate the need to employ more than one tool to evaluate the performance of deconvolution algorithms because there are many points of view concerning the tasks for which the reconstructions will be used.}

When the noise level increases, differences in SR between the PSFs are not as important as the noise level itself. The performances of MFBD and MISTRAL are more affected by noise (Fig. \ref{fig90}, top and middle row). In general, we found that IDAC has poor noise control. The criteria used by IDAC were not able to create a well-balanced trade-off between noise reconstruction and the achieved resolution enhancement, yielding either excessively noisy reconstructions or poor resolution. {When comparing AWMLE/ACMLE and MISTRAL in these noisier cases we see that the former still give better results when the SR mismatch is smaller than 10\%. These improvements are of the order of $2$-$4$dB in terms of PSNR or $10$-$20$\% in terms of MSSIM, depending on the noise level (readout noise of $\sigma=10$ or both shot and readout noise) and the mismatching between the PSFs (SR=53\% or SR=45\% cases).}

Figures \ref{fig110}, \ref{fig120} and \ref{fig130} show reconstructions, PSNR and SSIM maps for the SR=45\% calibrator. AWMLE, ACMLE and MFBD exhibit the typical ringing effect associated with strong transitions or edges (very visible at the limits of Saturn's body) whereas MISTRAL's $\mathit{L}_2-\mathit{L}_1$ edge-preserving prior attenuates such effects considerably. On the other hand, this prior must be responsible for the excessive attenuation of some elongated features present on Saturn's inner ring (Fig. \ref{fig110}, subpanels in row 4), while they have been detected by MFBD, AWMLE and ACMLE (although over-reconstructed in the latter two cases). We stress the fact that such features are not visible in the blurred and corrupted image (Fig. \ref{fig110}, subpanel 4.A). AWMLE and ACMLE exhibit the best results in the planet's body (in terms of the achieved resolution and noise attenuation), and also in the background space between Saturn's body and the innermost ring, which is a measurement of the algorithm's ability to suppress the noise (Figs. \ref{fig110}, \ref{fig120} and \ref{fig130}, subpanels in row 1). This effect is especially visible in Cassini's division (Figs. \ref{fig110}, \ref{fig120} and \ref{fig130}, subpanels in row 4). {We point out the curious opposite performance of MFBD and static-PSF approaches within Saturn's body: while MFBD achieves its better photometric result in the south pole, AWMLE and ACMLE obtain their better values in the north pole} (Fig. \ref{fig110}, subpanels 2.B, 2.D and 2.E). 

Amplification of noisy structures degrades the quality of PSNR, and especially, SSIM maps obtained by IDAC, probably to a lesser extent if compared to the typical Richardson-Lucy deconvolution (Fig. \ref{fig60} subpanel 2.A) but noise reconstruction is absolutely better controlled by the multiresolution support (Columns D and E in Figs. \ref{fig110}, \ref{fig120} and \ref{fig130}). MISTRAL also achieves good noise control. However, its performance is not better than that of AWMLE and ACMLE in the body part of Saturn (Fig. \ref{fig120}, subpanels 2,3.C vs. 2,3.D/E and Fig. \ref{fig130}, subpanels 2,3.C vs. 2,3.D/E), whereas MISTRAL's performance is more homogeneous across the rings (Fig. \ref{fig120}, subpanels 6.C vs. 6.D/E and Fig. \ref{fig130}, subpanels 6.C vs. 6.D/E) where AWMLE and ACMLE show poor behaviour, especially when both shot and readout noises are present. {On the other hand, photometric accuracy of MISTRAL is worse in the same region} (Fig. \ref{fig120}, subpanels 5.C vs. 5.D/E and Fig. \ref{fig130}, subpanels 5.C vs. 5.D/E), {thus showing again the complementarity between MSE and SSIM metrics. Finally, ACMLE has been able to reconstruct lines and elongated features within Saturn's body, at high noise levels} (Fig. \ref{fig120} and \ref{fig130}, subpanel 1.E) {without the visible dotted texture generated by the rest of algorithms, typical of non-complete elimination of noise.}

\section{CONCLUSIONS}

We have introduced a way of using multiresolution support, applied in the wavelet and curvelet domain, in the post-processing of adaptive-optics images. We have shown how a correlation-based mask applied over two simultaneously deconvolved images can help control the process of noise amplification. Furthermore, we have introduced the use of the structural similarity index as a measure of performance of different deconvolution algorithms (although it is not our aim in this paper to substitute the necessary subjective visual inspection of the results), {and compare its results with those achieved by the more common metric, the mean squared error, showing the need of using more than one metric for comparing the quality of deconvolved images}.
 
We have shown that the use of probabilistic masks can control noise amplification in a Richardson-Lucy scheme but, on the other hand, it does not prevent such amplification indefinitely (Fig. \ref{fig50}). {We also have shown that curvelet transform (CT) performs better than wavelet transform (WT) when a $\sigma$-based mask is used} (Fig. \ref{fig85}). 

One of the most important goals of this research was to devise an objective check of the typical assumption (within the AO astronomical community) about the supposed poor performance of static-PSF approaches with respect to the blind/myopic methods. For the dataset we used, the Richardson-Lucy scheme, controlled over the wavelet domain by a correlation mask, {and over the curvelet domain by a $\sigma$-based mask}, provided very competitive performance against the more well-known approaches like MFBD or regularized deconvolution (MISTRAL). Specifically, in the low-noise scenario it yielded 10-15\% better results (in terms of MSSIM) than IDAC and MISTRAL for mismatching in the PSF of up to approximately 8\% in terms of the Strehl ratio (middle panel, right column in Fig. \ref{fig90}). For higher noise levels the results of AWMLE and ACMLE were still 10-15\% better than those of MISTRAL, again up to 8\% mismatch, and they were 30-40\% better than those of IDAC. {In terms of PSNR, the improvement is of $1.5$-$2$dB with respect to MISTRAL and $2$-$3$dB with respect to IDAC. On the other hand, performance of MISTRAL and IDAC was photometrically better when highly-mismatched PSF of SR=36\% was employed as calibrator.}

\begin{figure*}
   \centering 
    \rotatebox{90}{\includegraphics[width=23cm]{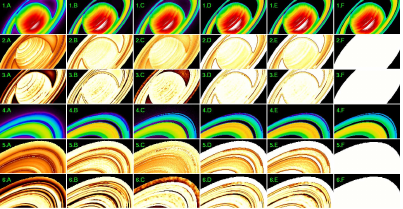}}
    {\caption{\footnotesize Row 1: Saturn's body; 1.A: Degraded and corrupted image with Gaussian noise ($\sigma=1$), 1.B: MFBD reconstruction (50 frames), 1.C: MISTRAL reconstruction, 1.D: AWMLE reconstruction (correlation-based mask), {1.E: ACMLE reconstruction ($\sigma$-based mask),} 1.F: Original image. {Row 2: corresponding PSNR maps.} Row 3: corresponding 4$\times$4 SSIM maps. Row 4: Saturn's rings detail. {Row 5: corresponding PSNR maps.} Row 6: corresponding 4$\times$4 SSIM maps. Reconstructions were performed with reference PSF at SR=45\%. Rows 1 and 4 are represented on a linear scale from 0 to 975, matching the dynamic range of the original image, i.e., subpanels 1.F/4.F. {Rows 2 and 5 are represented on a linear scale from 0 to 50.} Rows 3 and 6 are represented on a linear scale from 0 to 1. In rows 2, 3, 5 and 6 higher brightness indicates better quality.}}
  \label{fig110}
\end{figure*}

\begin{figure*}
   \centering 
    \rotatebox{90}{\includegraphics[width=23cm]{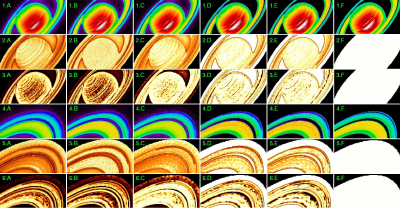} }
    \caption{\footnotesize  Row 1: Saturn's body; 1.A: Degraded and corrupted image with Gaussian noise ($\sigma=10$), 1.B: MFBD reconstruction (50 frames), 1.C: MISTRAL reconstruction, 1.D: AWMLE reconstruction (correlation-based mask), {1.E: ACMLE reconstruction ($\sigma$-based mask),} 1.F: Original image. {Row 2: corresponding PSNR maps.} Row 3: corresponding 4$\times$4 SSIM maps. Row 4: Saturn's rings detail. {Row 5: corresponding PSNR maps.} Row 6: corresponding 4$\times$4 SSIM maps. Reconstructions were performed with reference PSF at SR=45\%. Rows 1 and 4 are represented on a linear scale from 0 to 975, matching the dynamic range of the original image, i.e., subpanels 1.F/4.F. {Rows 2 and 5 are represented on a linear scale from 0 to 50.} Rows 3 and 6 are represented on a linear scale from 0 to 1. In rows 2, 3, 5 and 6 higher brightness indicates better quality.}
  \label{fig120}
\end{figure*}

\begin{figure*}
   \centering 
    \rotatebox{90}{\includegraphics[width=23cm]{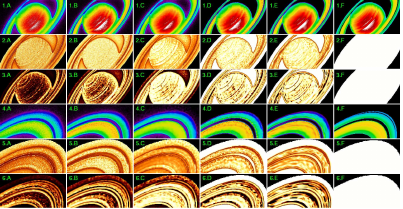}}
    \caption{\footnotesize  Row 1: Saturn's body; 1.A: Degraded and corrupted image with both shot and Gaussian noise ($\sigma=10$), 1.B: MFBD reconstruction (50 frames), 1.C: MISTRAL reconstruction, 1.D: AWMLE reconstruction (correlation-based mask), {1.E: ACMLE reconstruction ($\sigma$-based mask),} 1.F: Original image. {Row 2: corresponding PSNR maps.} Row 3: corresponding 4$\times$4 SSIM maps. Row 4: Saturn's rings detail. {Row 5: corresponding PSNR maps.} Row 6: corresponding 4$\times$4 SSIM maps. Reconstructions were performed with reference PSF at SR=45\%. Rows 1 and 4 are represented on a linear scale from 0 to 975, matching the dynamic range of the original image, i.e., subpanels 1.F/4.F. {Rows 2 and 5 are represented on a linear scale from 0 to 50.} Rows 3 and 6 are represented on a linear scale from 0 to 1. In rows 2, 3, 5 and 6 higher brightness indicates better quality.}
  \label{fig130}
\end{figure*}

The observed poor performance of IDAC, which can be visually appreciated in Figures \ref{fig110}-\ref{fig130}, can be explained in the following way. Multi-frame approaches rely on diverse PSFs to separate the blurring kernel from the object and to make the blind problem more tractable (less under-determined). Having more images helps when one has truly diverse PSFs. But this is not always the case with AO imaging. In fact, AO will stabilize the PSF and no number of new frames can then supply new information for MFBD. The standard deviation of the Strehl ratio value across the 200 frames used as input for IDAC was only 2\%. For all datasets that we had (10 stars) standard deviation across 200 frames rarely exceeded 5\%. It seems that single-image codes like MISTRAL or AW(C)MLE have an advantage in the case of stable AO observations.

Several lines of research are still open. It would be possible to study other types of masks, such as those based on the quadratic distance between related zones in consecutive wavelet planes \citep{Starck1994} or image segmentation (or wavelet plane segmentation) by means of neural networks in order to link and classify significant zones in the image \citep{Nunez1998}. Furthermore, {the use of other wavelet transforms, such as the undecimated Mallat trasform with three directions per scale,} and many other multitransforms should be studied and compared, e.g., shearlets \citep{Guo2007} or waveatoms \citep{Demanet2007}, to name only two.

Much more important, in our opinion, is studying how to include the multitransform support or probabilistic masks into blind and myopic approaches in order to improve their performance in terms of controlling the noise reconstruction inherent to every image reconstruction algorithm. This will be the topic of our future research.

\begin{acknowledgements}
We would like to thank Nicholas Law (University of Toronto) for supplying the PSFs from the Palomar Observatory. We extend our gratitude to Xavier Otazu (Universitat Aut\`onoma de Barcelona), Julian Christou (Gemini Observatory) and Laurent Mugnier (ONERA - The French Aerospace Lab) for useful comments and suggestions about the correct usage of algorithms AWMLE, IDAC and MISTRAL, respectively. We would also like to thank our referee whose suggestions and comments helped us to improve the paper. Effort sponsored by the Air Force Office of Scientific Research, Air Force Material Command (USAF) under grant number FA8655-12-1-2115, and by the Spanish Ministry of Science and Technology under grant AyA-2008-01225. The U.S Government is authorized to reproduce and distribute reprints for Governmental purpose notwithstanding any copyright notation thereon.
\end{acknowledgements}

\bibliographystyle{aa}

\end{document}